\documentclass[aps,showpacs]{revtex4}%
\usepackage{amssymb}
\usepackage{amsmath}
\usepackage{graphicx}
\usepackage{amsfonts}%
\begin{document}
\preprint{HEP/123-qed}
\title[Synchrotron Radiation in Myers-Pospelov electrodynamics]
{Synchrotron Radiation in Lorentz-violating Electrodynamics: 
\\the Myers-Pospelov model}
\author{R. Montemayor$^{1}$ and L.F. Urrutia$^{2}$}
\affiliation{1. Instituto Balseiro and CAB, Universidad Nacional
de Cuyo and CNEA, 8400 Bariloche, Argentina\\
2. Departamento de F\'\i sica de Altas Energ\'\i as, Instituto de
Ciencias Nucleares, Universidad Nacional Aut\'onoma de M\'exico,
Apartado Postal 70-543, 04510 M\'exico D.F.}

\begin{abstract}
We develop a detailed analysis of synchrotron radiation in the effective
Lorentz invariance violating (LIV) model of Myers-Pospelov, considering
explicitly both the dynamics of the charge producing the radiation and the
dynamics of the electromagnetic field itself. Within the radiation
approximation we compute exact expressions in the LIV parameters for the
electric and magnetic fields, the angular distribution of the power spectrum,
the total emitted power in the m-th harmonic and the polarization. We also
perform expansions of the exact results in terms of the LIV parameters to
identify the dominant effects, and study the main features of the high energy
limit of the spectrum. A very interesting consequence is the appearance of
rather unexpected and large amplifying factors associated with the LIV
effects, which go along with the usual contributions of the expansion
parameter. This opens up the possibility of looking for astrophysical sources
where these amplifying factors are important to further explore the
constraints imposed upon the LIV parameters by synchrotron radiation
measurements. We briefly sketch some phenomenological applications in the case
of SNRs and GRBs.

\end{abstract}
\maketitle


\section{Introduction}

The isotropy together with the Lorentz covariance of the laws of
nature in locally inertial frames are properties which must be
ultimately determined on experimental grounds. This is
particularly true for boosts which, due to the non-compact
character of the Lorentz group, can always be tested at higher
energies. The experimental study of these questions began around
1960 with the works of  Hughes et. al. and Drever \cite{HUDREV}
and has continued since then involving a huge variety of very
different testing methods with ever increasing experimental
precision \cite{CPTMEETS}. It has been possible to correlate such
highly disparate observations in terms of the Standard Model
Extension (SME) proposed in Ref. \cite{KOSTELECKY}. This model is
an effective theory that parameterizes all the possible Lorentz
violating interactions consistent with the standard model, which
arise in a spontaneously broken formulation of such symmetry
violation motivated by string theory \cite{KOSTELECKY1}. The SME
has been recently generalized to incorporate gravity effects
\cite{KOSTELECKY2}.

A possible physical realization for such Lorentz invariance violations has
recently been suggested within the context of dynamical modifications induced
by quantum gravity in the propagation of particles in flat space, as for
example in Ref. \cite{ACNAT}. Most theories of non-perturbative quantum
gravity predict that our notion of space as a continuum manifold has to be
abandoned at distances of the order of the Planck length $\ell_{P}
\approx1/M_{P}\approx10^{-33}\,cm$, where space should be regarded
as a bunch of quantum excitations. A first question related to
this idea is whether or not such a drastic change in the
description could produce observable residual modifications to our
large scale description of particle dynamics, manifested as
additional terms in the corresponding actions that are suppressed
by powers of $\ell_{P}$. Some approaches suggest that standard
Lorentz covariance is consistent with the granular structure of
space predicted by quantum gravity \cite{ROVSPE,SORKIN}, expressed
as the absence of $\ell_{P}$-corrections to large scale dynamics,
while others propose that such corrections should arise, but with
different interpretations of their manifestations. The most direct
interpretation of such corrections is through the existence of a
preferred reference frame associated to the space granularity in
which the existence of a minimum length provides a maximum
physical upper bound for the momenta accessible in such a frame.
Nevertheless, in Ref. \cite{CPSUV} it was shown that this
interpretation leads to severe fine-tuning problems when
considering the radiative corrections induced by the standard
interactions. Supersymmetry has been proposed as a way to improve
such situation \cite{NIBBPOS}. An alternative proposal can be
found in Ref. \cite{ALFARO05}. A second possibility arises when
considering the dynamical modifications as signaling a spontaneous
breaking of the Lorentz symmetry. In this case the corresponding
effective theory also possesses a class of preferred frames that
are those where the chosen vacuum expectation values of the
respective fields are only a very small perturbation with respect
to the Lorentz symmetric case. These are the so called concordant
frames \cite{LEHNERT} and they are not directly related to the
idea of a maximum momentum. The SME together with effective models
like those of Refs. \cite{NIBBPOS,MP} belong to this category. A
third possibility, generically included in the so called deformed
special relativity approach, allows for dynamical corrections
without the introduction of preferred reference frames by
generalizing the concept of relative inertial frames on the basis
of non-linear representations of the Poincar\'{e} group
\cite{DSRAC,MASMOL,VISSER,GIRLIV,SMOLINR}. Finally, alternative
ways of incorporating such dynamical modifications can be found in
Ref. \cite{ALTERN}. For recent reviews of the above topics see for
example Ref. \cite{REVIEW1,REVIEW2,REVIEW3}.

One of the most studied consequences of the Lorentz violating corrections to
the dynamics are related with multiple phenomena associated to the modified
dispersion relations (MDR) for photons and fermions
\begin{equation}
\omega^{2}(k)=k^{2}\;\pm\xi\frac{k^{3}}{M},\quad E^{2}(p)=p^{2}+m^{2}%
+\eta_{R,L}\frac{p^{3}}{M},\label{DR}%
\end{equation}
respectively. Here $M$ is a scale that signals the onset of LIV
and is usually associated with the Planck mass $M_{P}$. In this
context the main emphasis has been set on the determination of
observational bounds for the LIV parameters $\xi$ and $\eta_{R,L}$
appearing in the MDR, thus opening the door to quantum gravity
phenomenology \cite{ACQGPHEN}. A partial list of references is
given in \cite{analysis,JACOBSON1,JACOBSON3,JACOBSON2}.

The possibility of deriving such modified dynamics from a more fundamental
theory has been further explored in Loop Quantum Gravity, where constructions
of increasing degree of sophistication have managed to produce modified
effective actions for photons and fermions encoding the relations (\ref{DR}).
Basically, the starting point of these approaches has been the well defined
Hamiltonian operators of the quantum theory, together with a heuristical
characterization of the semiclassical ground state. The first derivation of
such a type of dispersion relations in the context of Loop Quantum Gravity was
developed in Ref. \cite{GPED}, and leads to a consistent extension of Maxwell
electrodynamics with correction terms linear in $\ell_{P}$. Subsequently, an
alternative approach inspired by Thiemann's regularization \cite{THIEM1}
reproduced and extended such a result for the photon case \cite{AMU1} and
produced the corresponding modified Hamiltonian for the fermion case
\cite{AMU2}. For a review see for example Ref. \cite{REVURRU}. Coherent states
\cite{THIEMANN} have also been constructed, leading to similar types of
dynamical modifications. All previous approximations have been kinematical,
i.e. without incorporating the dynamical constraint of general relativity. In
other words, a physically sound characterization of the semiclassical ground
state corresponding to a given asymptotic classical metric that dictates the
appropriate symmetries of the effective low energy theory is still lacking.
This issue remains one of the most important open problems in loop quantum
gravity. String theory has also provided a possible connection between quantum
gravity and LIV in the case of photons and fermions \cite{string,ELLIS1}.

 Let us make some very general comments regarding some of the
bounds already found in the literature for the LIV parameters
appearing in  MDR  of the form given in Eq. (\ref{DR}) for
photons, electrons \ and protons, arising from dimension five
corrections to standard electrodynamics. The main scenarios that
we consider here are phenomena related to observations of
electromagnetic radiation arising from astrophysical objects and
from the arrival of ultra high energy cosmic rays (UHECR).

The absence of birefringence effects from distant galaxies
provides the bound $\left\vert \xi \right\vert \lesssim
10^{-4}$ \cite{Gleiser-Kozameh}. Observations from the Crab
nebula, a type of supernova remnant (SNR), give the following
bounds on the parameters. On one hand the birefringence
limit has been improved up to $\left\vert \xi \right\vert
\lessapprox 10^{-6}$ for the parameter region $|\eta
_{e}|<10^{-4}$ with $\eta _{e}$ being at least one of the $\eta
_{L,R}$ for electrons or positrons \cite{REVIEW3}. On the
other hand, the intersection of the corresponding synchrotron,
synchrotron-Cerenkov and the inverse Compton constraints produces
the two-sided upper bound $|\eta _{L,R}|<10^{-2}$ on one of the
two parameters $\eta _{L,R}$ \cite{REVIEW3}. Typical upper bounds
for the corresponding energies in the Crab nebula are:
$10^{6}\;GeV$ for the radiating electrons (positrons),
$10^{-1}\;GeV$ for the synchrotron photons and $5\times
10^{4}\;GeV$\ for the inverse Compton photons. These give an idea
of the energy scale in which the above bounds are obtained.

 The assumption of a  dispersion relation of the type (\ref{DR})
with a universal LIV parameter $\eta$ for all particles in the
study of UHECR has produced the stringent bound $|\eta|< 10^{-32}
\, GeV^{-1}$ \cite{AP2003}. This reference also reports fits of
the UHECR spectrum starting from $10^{18} eV$ which include the
AGASA data.

The vacuum Cerenkov effect becomes significant for charged
particles with high energy. It has been already discussed, both
using a kinematical approach \cite{CG,JACOBSON3,GM}, and on the
basis of a dynamical analysis in the context of the
Maxwell-Chern-Simons action \cite{LP}. An analogous analysis could
be interesting in the Myers-Pospelov model, but it is out of the
scope of the present work. Very stringent boundaries for the LIV
parameters from the absence of vacuum Cerenkov radiation have been
recently presented using protons in the ultra high energy region
$E_{proton}\simeq 3\times 10^{11}GeV$, based on reasonable
assumptions about the nature of the high energy cosmic rays.
Assuming that the values for the LIV parameter $ \eta $ are
comparable for different standard model elementary species, or at
least the same within species of a given spin, leads to consider
unmodified dispersion relations for a structureless proton, i. e.
$\eta _{p}=0$. With these assumptions the bound
$|\tilde{\xi}|\lessapprox 2\alpha _{p}=\left(
4m_{proton}^{2}/E_{proton}^{3}\right) \simeq
10^{-34}\;GeV^{-1}$ is obtained in Ref. \cite{GM}. Allowing for
MDR for the proton, $\tilde{\eta}_{p}\neq 0$, the less stringent
bounds of Ref. \cite{JACOBSON3} are recovered:
$\tilde{\eta}_{p}<\alpha _{p}/4,\;\;\;|\tilde{\xi}%
|<\left\vert \tilde{\eta}_{p}-\alpha _{p}-\alpha _{p}\sqrt{1-4\tilde{\eta}%
_{p}/\alpha }\right\vert $. Here $\tilde{\eta}_{p}=\eta
_{p}/M_{P}$ and refers to the proton, while $\tilde{\xi}=\xi
/M_{P}$. These constraints apply to both photon helicities, which
is the origin for the bound on the absolute
value $|\tilde{\xi}|$. Considering the proton polarization
we can write $\tilde \eta = \lambda_p k_p$, where $\lambda_p=\pm 1
$ is the corresponding polarization. Thus, if we have
$k_p>\alpha_p/4$ ($k_p<-\alpha_p/4$) the protons with positive
(negative) helicity emit vacuum Cerenkov radiation, while the
protons with opposite helicity do not radiate. Hence, to
establish boundaries from Cerenkov radiation it is necessary not
only to know if there are protons at a given energy, but also
what polarizations they have. At energies $ E\simeq 10^{11}\
GeV$ the UHECR are detected by the cascade they produce in the
atmosphere, where the information about primary proton
polarization is lost. To determine their polarization they must
be observed directly as primary cosmic rays. The highest energy
for which this could be done is $E\simeq 10^{5}\ GeV$, with
instruments on board of satellites \cite{grieder}. In this case
$\alpha _{p}\simeq 10^{-15}\, GeV^{-1}$ and thus we would get
$|\tilde{\xi}|\lessapprox 10^{4}$, which is  $10^{10}$ times
weaker than the boundary obtained from vacuum birefringence. It
is interesting to remark that the highest electron energies
measured in primary cosmic rays are estimated to be of the order
of $ 10^{3}\;GeV$ with a flux 0.05 times that of the protons
\cite{grieder}. This gives $\alpha _{e}=10^{-15}\ GeV^{-1}$. In
the Crab nebula we have $\alpha _{e}\simeq 10^{-24}\;GeV^{-1}$ 
which produces a much better bound than electron vacuum Cerenkov
radiation in cosmic rays.

For a detailed and updated discussion of the LIV parameter
constraints see for example Refs. \cite{REVIEW1,REVIEW2,REVIEW3}.

Now let us go back to the observation of 100 MeV sinchrotron
radiation from the Crab nebula. This fact alone leads to the bound
$\pm \eta _{L,R}>-7\times 10^{-8}$ for at least one of the four
possibilities \cite{JACOBSON1}. By itself, this does not impose
any constraints on the parameters, but it plays a role when
combining it with the vacuum Cerenkov effect. Such a bound is
based on a set of very reasonable assumptions on how some of the
standard results of synchrotron radiation extend to the Lorentz
non-invariant situation. This certainly implies some dynamical
assumptions, besides the purely kinematical ones embodied in
(\ref{DR}). In this paper we examine these assumptions in the
light of a particular model, which we choose to be the classical
version of the Myers-Pospelov (MP) effective theory, that
parameterizes LIV using dimension five operators \cite{MP}.
Furthermore, a complete calculation of synchrotron radiation in
the context of this model is presented. This constitutes an
interesting problem on its own whose resolution will subsequently
allow the use of additional observational information to put
bounds upon the correction parameters. For example we have in mind
the polarization measurements from cosmological sources. The case
of gamma ray bursts (GRB) has recently become increasingly
relevant \cite{CB}, although it is still at a controversial stage
\cite{CONTR}. Our calculations rest heavily upon the work by
Schwinger et al. on synchrotron radiation, reported in Refs.
\cite{SCH49,SCHWBOOK,SCWANNP}. In this work we restrict ourselves
to classical theories, and therefore we do not address the
fine-tuning problems arising from quantum corrections associated
with the coexistence of space granularity and preferred reference
frames \cite{CPSUV}. A partial list of previous studies of
electrodynamics incorporating LIV via dimension three and four
operators is given in Refs. \cite{ALL,LP}. A summary of the
present work has already been reported in Ref. \cite{PLBMU}.

The paper is organized as follows. In Section II we obtain the
equations of motion for the charge and electromagnetic sectors of
the MP model. The polarized electromagnetic fields for arbitrary
sources in the radiation approximation are subsequently calculated
in Section III using the standard Green function approach. In
Section IV we obtain the general expression for the angular
distribution of the power spectrum. This is subsequently applied
in the next section to the case of the synchrotron radiation
produced by a charged particle moving in a plane perpendicular to
a constant magnetic field. The corresponding angular distribution
of the radiated power in each harmonic is obtained, together with
the corresponding total power. Also the associated Stokes
parameters are calculated in this section. The expansions of such
exact results to leading order in the LIV electromagnetic
parameter are contained in Section VI, where we highlight the
dominating effects of the Lorentz violation. A rough estimation
of the relative contributions arising from dimension six
operators is included in seccion VII. Motivated by the estimated
parameters of the astrophysical objects under consideration we
discuss the large harmonic expansion of the previously found
results in Section VIII. We analyze some phenomenological
consequences of our results in Section IX. Finally, Section X
contains a summary of the main results obtained in the paper
together with a discussion of accessible regions for the LIV
parameters in the cases of Crab Nebula, Mk 501, and the
$\gamma$-ray burst GRB021206. There are also two Appendices
containing frequently used material.

\section{The Myers-Pospelov electrodynamics}

The Myers-Pospelov approach is based on an effective field theory that
describes Lorentz violations generated by dimension five operators,
parameterized by the velocity $n^{\mu}$ of a preferred reference frame which
is not taken as a dynamical field. These operators are assumed to be
suppressed by a factor $M_{P}^{-1}$, and hence can be considered as small
perturbations at the classical level. In the following we analyze the
electromagnetic radiation of a classical charged particle in this framework.
As a first step in this approach we characterize the particle and the
electromagnetic field dynamics, introducing the charge-electromagnetic field
interaction via the minimal coupling, in such a way that the usual gauge
symmetry is maintained in this theory.

\subsection{Charged particle dynamics}

The dynamics of a classical charged particle can be obtained from the action
for a scalar charged field. In this case the Myers-Pospelov action is
\begin{equation}
S_{MP}=\int d^{4}x\;\left[  \partial_{\mu}\varphi^{\ast}\partial^{\mu}%
\varphi-\mu^{2}\varphi^{\ast}\varphi+i\tilde{\eta}\varphi^{\ast}\left(
V\cdot\partial\right)  ^{3}\varphi\right]  ,
\end{equation}
with the notation $V\cdot\partial=V^{\mu}\,\partial_{\mu}$. In momentum space,
where we write $\varphi(x)=\varphi_{0}\;\exp i(p^{0}t-\mathbf{p\cdot x})$, and
in the reference frame where $V^{\alpha}=\left(  1,\mathbf{0 }\right)  $, the
dispersion relation becomes
\begin{equation}
\left(  p^{0}\right)  ^{2}+\tilde{\eta}\left(  p^{0}\right)  ^{3}=\mathbf{p}
^{2}+\mu^{2} . \label{EXDR}%
\end{equation}
To compare with Jacobson et al. results
\cite{JACOBSON1,JACOBSON3,JACOBSON2}, the parameter $\tilde{\eta}$
can be written
\begin{equation}
\tilde{\eta}=-\frac{\eta}{M_{P}},\;\;\;\eta<0,
\end{equation}
where $M_{P}$ is the Planck mass and $\eta$ is a dimensionless constant. The
equation (\ref{EXDR}) is an exact relation in $\tilde{\eta}$. From here we
obtain the Hamiltonian for a massive particle to second order in $\tilde{\eta
}$
\begin{equation}
p^{0}=H=\left(  \mathbf{p}^{2}+\mu^{2}\right)  ^{1/2}-\frac{1}{2}\tilde{\eta
}\left(  \mathbf{p}^{2}+\mu^{2}\right)  +\frac{5}{8}\tilde{\eta}^{2}\left(
\mathbf{p} ^{2}+\mu^{2}\right)  ^{3/2}+O(\tilde{\eta}^{3}).
\end{equation}
In the following we will consider the interaction of a particle of mass $\mu$
and charge $q$ with a static magnetic field ($\phi=0,\;\mathbf{A}(\mathbf{r}%
)$). The standard minimal coupling produces the Hamiltonian%
\begin{equation}
H=\left[  \left(  \mathbf{p-}\frac{q}{c}\mathbf{A}\right)  ^{2}+\mu
^{2}\right]  ^{1/2}-\frac{1}{2}\tilde{\eta}\left[  \left(  \mathbf{p-}\frac
{q}{c}\mathbf{A} \right)  ^{2}+\mu^{2}\right]  +\frac{5}{8}\tilde{\eta}%
^{2}\left[  \left(  \mathbf{p-}\frac{q}{c}\mathbf{A}\right)  ^{2}+\mu
^{2}\right]  ^{3/2}+O(\tilde{ \eta}^{3}).
\end{equation}
Here $c=3\times10^{10}\, cm\, s^{-1}$ denotes the uncorrected velocity of
light in vacuum. In the sequel we set $c=1$. Observe that the dispersion
relation (\ref{EXDR}) provides the exact inversion
\begin{equation}
\left(  \mathbf{p}-q\mathbf{A}\right)  ^{2}=\left(  1+\tilde{\eta} E\right)
E^{2}-\mu^{2}, \label{PMA}%
\end{equation}
with $E$ being the energy of the particle, which together with the Hamilton
equations yield
\begin{align}
\dot{x}_{i}  &  =v_{i}=\frac{\partial H}{\partial p_{i}}=\frac{1}{E}\left(  1-
\frac{3}{2}\tilde{\eta}E+\frac{9}{4}\tilde{\eta}^{2}E^{2}\right)  \left(
p_{i}-q A_{i}\right)  ,\label{VEL}\\
\dot{p}_{i}  &  =-\frac{\partial H}{\partial x_{i}}=q\dot{x}_{j}
\frac{\partial A_{j}}{\partial x_{i}},
\end{align}
up to second order in $\tilde{\eta}$. The acceleration results
\begin{equation}
\mathbf{\ddot{r}}=\frac{q}{E}\left(  1-\frac{3}{2}\tilde{\eta}E+\frac{9}{4}
\tilde{\eta}^{2}E^{2}\allowbreak\right)  \left(  \mathbf{v\times B}\right)  .
\end{equation}
As in the usual case, this means that the magnitude $|\mathbf{v|}$ is constant
and that the projection of the orbit in a plane orthogonal to
$\mathbf{B=\nabla\times A}$ is a circular orbit with a Larmor frequency
\begin{equation}
\omega_{0}=\frac{|q|B}{E}\left(  1-\frac{3}{2}\tilde{\eta}E+\frac{9}{4}
\tilde{\eta}^{2}E^{2}\allowbreak\right)  . \label{OMEGA0}%
\end{equation}
In general the motion is a helix with pitch angle (the angle between the
velocity and the magnetic field) $\alpha$. Let us define
\begin{equation}
\beta_{\perp}=\beta\sin\alpha,\qquad\beta_{\parallel}=\beta\cos\alpha,
\end{equation}
where we emphasize that we are using the standard definition\ $\beta=|
\mathbf{v|}/c$. The solution to the equations of motion can be written as
\begin{equation}
\mathbf{r(}t)\mathbf{=}\left(  \frac{\beta_{\perp}}{\omega_{0}}\cos\omega
_{0}t,\ \frac{\beta_{\perp}}{\omega_{0}}\sin\omega_{0}t,\ \beta_{\parallel
}t\right)  , \label{rt}%
\end{equation}
and hence
\begin{equation}
\mathbf{v(}t)\mathbf{=}(-\beta_{\perp}\sin\omega_{0}t,\;\beta_{\perp}%
\cos\omega_{0}t,\;\beta_{\parallel}). \label{vt}%
\end{equation}
The orbit equation (\ref{rt}) identifies
\begin{equation}
R=\frac{\beta_{\perp}}{\omega_{0}} \label{RADIOUS}%
\end{equation}
as the radius of the projection of the helix in the plane perpendicular to
$\mathbf{B}$. From Eqs. (\ref{PMA}) and (\ref{VEL}) we obtain
\begin{equation}
1-\beta^{2}=\frac{\mu^{2}}{E^{2}}\left[  1+2\frac{\tilde{\eta}E^{3}}{\mu^{2}%
}-\frac{15}{4}\frac{\tilde{\eta}^{2}E^{4}}{\mu^{2}}+O(\tilde{\eta}^{3})
\right]  , \label{UMB2}%
\end{equation}
where the considered range of energies is such that
\begin{equation}
\frac{\mu}{E}<<1,\;\;\;\;\;\tilde{\eta}E<<1. \label{SMALLQ}%
\end{equation}
According to the relation among the quantities appearing in Eq.
(\ref{SMALLQ}), we can consider different energy ranges to write
approximate expressions for the Lorentz factor
\begin{equation}
\gamma=\left(  1-\beta^{2}\right)  ^{-1/2}. \label{LORFACT}%
\end{equation}
In Eq. (\ref{UMB2}), the ratio between the second order term in $\tilde{\eta}
$ and the first order one is proportional to $\tilde{\eta}E\ll1$. Therefore,
assuming $\tilde{\eta}\gtrsim10^{-19}\ GeV^{-1}$, the last term is negligible
compared with the second one at least for $E\lesssim10^{19}\ GeV$. The last
term is of order one when $\tilde{\eta}^{2}E^{4}/\mu^{2}\simeq1$, i.e. when
$E\simeq\left(  \mu/\tilde{\eta}\right)  ^{1/2}$. For electrons this means
$E\simeq7\times10^{7}\ GeV\simeq7\times10^{4}\ TeV$, and an even larger energy
for more massive particles. Therefore, in the range of energies expected in
astrophysical objects and when $E<7\times10^{7}\ GeV$, the $\gamma$ factor can
be written as
\begin{equation}
\gamma\simeq\frac{E}{\mu}\left(  1+2\frac{\tilde{\eta}E^{3}}{\mu^{2}} \right)
^{-1/2}. \label{18}%
\end{equation}
Hence we have the two possible approximate expressions
\begin{align}
\gamma &  \simeq\frac{E}{\mu}\left(  1+\frac{\tilde{\eta}E^{3}}{\mu^{2}}
\right)  ,\ \ \ \ \ \ \ \ \ \qquad\frac{\tilde{\eta}E^{3}}{\mu^{2}}%
\ll1,\label{g1}\\
\gamma &  \simeq\sqrt{\frac{1}{2\kappa E}}\left(  1-\frac{\mu^{2}}{4\tilde{
\eta}E^{3}}\right)  ,\qquad\frac{\tilde{\eta}E^{3}}{\mu^{2}}\gg1. \label{g2}%
\end{align}
The limiting condition between these two energy ranges is $E\simeq10^{4}\ GeV$
for electrons and $E\simeq10^{6}\, GeV$ for protons. In the case of electrons
in a SNR of the type of Crab Nebula, the maximum energy of the electrons is of
the order of $10^{3}\,TeV$, and therefore the approximation (\ref{g1}) still holds.

According to the preceding analysis, the current for a charged particle moving
in the general helical motion is
\begin{equation}
\mathbf{j}(t,\mathbf{r})=q\delta^{3}(\mathbf{r}-\mathbf{r}(t))\,\mathbf{v}(t),
\label{CURRENT}%
\end{equation}
where $\mathbf{r}(t)$ and $\mathbf{v}(t)$ are given in Eqs. (\ref{rt}) and
(\ref{vt}) respectively, with $\beta$ determined by the Lorentz factor
(\ref{g1}). In the following we will consider only circular motion, i.e.
$\alpha=\pi/2$.

\subsection{The electromagnetic field}

The Myers-Pospelov action for the electromagnetic field is%
\begin{equation}
S_{MP}=\int d^{4}x\left[  -\frac{1}{4}F_{\mu\nu}F^{\mu\nu}-4\pi\,j^{\mu}%
A_{\mu}+\tilde{\xi}\left(  V^{\alpha}F_{\alpha\delta}\right)  (V\cdot
\partial)(V_{\beta}\tilde{F}^{\beta\delta})\right]  , \label{mpa}%
\end{equation}
where $\tilde{\xi}=\xi/M_{P}$, with $\xi$ being a dimensionless parameter. We
are using the conventions
\begin{equation}
\;\tilde{F}^{\beta\delta}=\frac{1}{2}\epsilon^{\beta\delta\rho\sigma}%
F_{\rho\sigma},\;\;\;\;\epsilon^{0123}=+1,\;\;\;\;\epsilon_{123}%
=+1,\;\;\;\;\eta=(+,-,-,-),\;\;\;\epsilon^{0ijk}=\epsilon_{ijk}.
\end{equation}
As usual we have $F_{\mu\nu}=\partial_{\mu}A_{\nu}-\partial_{\nu}A_{\mu}$ and
$j^{\mu}=(\rho,\mathbf{j})$, with the electric and magnetic fields identified
according to
\begin{align}
F_{0i}  &  =E_{i},\;\;\;F_{ij}=-\epsilon_{ijk}B_{k},\;\;\;\;B_{k}=-\frac{1}%
{2}\epsilon_{kij}F_{ij},\label{CONV1}\\
\tilde{F}_{0i}  &  =B_{i},\;\;\;\tilde{F}_{ij}=\epsilon_{ijk}E_{k}%
,\;\;\ \ \;\;E_{k}=\frac{1}{2}\epsilon_{kij}\tilde{F}_{ij}. \label{CONV2}%
\end{align}
These lead to the standard homogeneous Maxwell equations
\begin{equation}
\nabla\cdot\mathbf{B}=0\mathbf{,\;\;\;\nabla\times E}+\frac{\partial
\mathbf{B}}{\partial t}=0.
\end{equation}
The equations of motion obtained from the action (\ref{mpa}) are
\begin{equation}
\partial_{\mu}F^{\mu\nu}+\tilde{\xi}(V\cdot\partial)\left(  \epsilon^{\beta
\nu\delta\rho}V_{\beta}\left(  V^{\alpha}\partial_{\rho}F_{\alpha\delta
}\right)  -\left(  V^{\alpha}\partial_{\alpha}\right)  (V_{\beta}\tilde
{F}^{\beta\nu})\right)  =4\pi\,j^{\nu}. \label{em}%
\end{equation}
For simplicity, we keep working in the rest frame $V^{\alpha}=(1,
\mathbf{0})$. If required we can boost to any other reference
frame with a given $V^{\alpha}$ by means of an observer (passive)
Lorentz transformation. Rewriting Eqs. (\ref{em}) in vector form
we get
\begin{align}
\nabla\cdot\mathbf{E}  &  =4\pi\rho,\\
-\frac{\partial\mathbf{E}}{\partial t}+\nabla\times\mathbf{B}+\tilde{\xi}%
\frac{\partial}{\partial t}\left(  -\nabla\times\mathbf{E}+\frac
{\partial\,\mathbf{B}}{\partial t}\right)   &  =4\pi\mathbf{j.}%
\end{align}
Thus, in terms of the standard potential fields $A^{\mu}=\left(
\phi,\mathbf{A} \right)  $ we are left with
\begin{align}
\nabla^{2}\phi+\frac{\partial}{\partial t}\nabla\cdot\mathbf{A}  &  =-4\pi
\rho,\\
\frac{\partial}{\partial t}\left(  \nabla\phi+\frac{\partial\mathbf{A}%
}{\partial t}\right)  +\nabla\times\left(  \nabla\times\mathbf{A}\right)
+2\tilde{\xi}\nabla\times\frac{\partial^{2}\,\mathbf{A}}{\partial t^{2}}\,  &
=4\pi\,\mathbf{j},
\end{align}
which can be conveniently expressed in the radiation (Coulomb) gauge,
$\nabla\cdot\mathbf{A}=0$, as
\begin{align}
\phi &  =-4\pi\frac{1}{\nabla^{2}}\rho,\\
\frac{\partial^{2}\mathbf{A}}{\partial t^{2}}-\nabla^{2}\mathbf{A}+2\tilde
{\xi}\nabla\times\frac{\partial^{2}\mathbf{A}}{\partial t^{2}}  &
=4\pi\,\left(  \mathbf{j-}\nabla\frac{1}{\nabla^{2}}\nabla\cdot\mathbf{j}%
\right)  \equiv4\pi\,\mathbf{j}_{T}, \label{ema}%
\end{align}
where the electric and magnetic fields reduce to
\begin{equation}
\mathbf{E}=-\frac{\partial\mathbf{A}}{\partial t},\;\;\;\;\mathbf{B}%
=\nabla\times\mathbf{A},
\end{equation}
in the corresponding large distance approximation.

The energy momentum tensor $T_{\mu\nu}$ for this modified electrodynamics is
given by%
\begin{align}
T_{\;0}^{0}  &  =\frac{1}{8\pi}(\mathbf{E}^{2}+\mathbf{B}^{2})-\frac
{\tilde{\xi}}{4\pi}\mathbf{E\,\cdot\,}\frac{\partial\mathbf{B}}{\partial
t},\label{ed}\\
\mathbf{S}  &  =\frac{1}{4\pi}\mathbf{E}\times\mathbf{B}-\frac{\tilde{\xi}%
}{4\pi}\mathbf{E}\times\frac{\partial\mathbf{E}}{\partial t}, \label{pv}%
\end{align}
which are exact expressions in $\tilde{\xi}$. These components satisfy the
usual conservation equation
\begin{equation}
\frac{\partial T^{00}}{\partial t}+\mathbf{\nabla}\cdot\mathbf{S}=0,
\end{equation}
outside the sources.

To solve the equation of motion for $\mathbf{A}$, Eq. (\ref{ema}), we can go
to the momentum space with the convention
\begin{equation}
F(t,\mathbf{r})=\int\frac{d^{4}k}{\left(  2\pi\right)  ^{4}}e^{-i\omega
t+i\mathbf{k}\cdot\mathbf{r}}F(\omega,\mathbf{k}).
\end{equation}
The different types of Fourier transforms are specified by the corresponding
arguments. For example, if $F(t,\mathbf{r)}$ denotes the function in
space-time, $F(\omega,\mathbf{r)}$ denotes the Fourier transformed function to
frequency space, while $F(\omega,\mathbf{k})$ denotes the Fourier transformed
function to frequency and momentum spaces. In this way Eq. (\ref{ema}) reduces
to
\begin{equation}
\left(  -\omega^{2}+k^{2}-2i\tilde{\xi}\omega^{2}\,\mathbf{k}\,\times\right)
\mathbf{A(}\omega\mathbf{,k)}=4\pi\,\mathbf{j}_{T}\mathbf{(}\omega
,\mathbf{k)}.
\end{equation}
This equation can be diagonalized using the basis of circular polarization
$\lambda=\pm1$, defined in Appendix A, giving
\begin{equation}
\left(  -\omega^{2}+k^{2}\mp2\tilde{\xi}\omega^{2}k\right)  \mathbf{A}^{\pm
}\mathbf{(}\omega\mathbf{,k)}=4\pi\mathbf{j}_{T}^{\pm}\mathbf{(}%
\omega\mathbf{,k)}. \label{DEFPOL}%
\end{equation}

\section{Green functions and fields}

The simplest way to proceed is by introducing the retarded Green functions
with definite polarization
\begin{equation}
G_{ret}^{\lambda}(\omega,\mathbf{k})=\left.  \frac{1}{ k^{2}-\lambda
2\tilde{\xi}\omega^{2}k-\omega^{2} }\right\vert _{\omega\rightarrow
\omega+i\epsilon}, \label{GRETGF0}%
\end{equation}
together with the total retarded Green function $ [G_{ret}(\omega
,\mathbf{k})]_{ik}=\sum_{\lambda}\, P_{ik}^{\lambda
}\,G_{ret}^{\lambda}(\omega,\mathbf{k}) $ and to calculate
$[G_{ret}(\omega , {\vec r}- {\vec r}^{\prime })]_{ik}$. Here
$P_{ik}^{\lambda }$ is the helicity projector defined in
Eq. (\ref{PROY}). The angular integration can be immediately
performed and the remaining integration over $dk$ gets
contributions only from the upper half plane in the complex
variable $k$. Finally one can identify the polarization
components of the total Green function as
\begin{equation}
G_{ret}^{\lambda}\mathbf{(}\omega\mathbf{,r-r}^{\prime}\mathbf{)=}\int
\frac{d^{3}\mathbf{k}}{\left(  2\pi\right)  ^{3}}e^{i\mathbf{k}\cdot
(\mathbf{r-r}^{\prime})}G_{ret}^{\lambda}\mathbf{(}\omega\mathbf{,k)=}\frac
{1}{4\pi R}\frac{n(\lambda z)}{\sqrt{1+z^{2}}}e^{in(\lambda z)\omega R},
\label{GRETGF}%
\end{equation}
where $R=|\mathbf{r}-\mathbf{r}^{\prime}|$.  Here we introduce the
polarization-dependent refraction index $n(\lambda z)$
\begin{equation}
n(\lambda z)=\sqrt{1+z^{2}}+\lambda z,\;\;\mathbf{\ }z=\tilde{\xi}\omega.
\label{REFIND}%
\end{equation}
In this way, the fields $\mathbf{A^{\lambda}}$ in Eq. (\ref{DEFPOL}) have well
defined phase velocities $v_{\lambda}=1/n(\lambda z)$ and this situation can
be described as the propagation of photons in a dispersive birefringent medium.

The Green functions (\ref{GRETGF}) determine the corresponding potentials with
the standard replacements $1/R\simeq1/| \mathbf{r}|\equiv1/r$ in the
denominator and $R\simeq r-\mathbf{\hat{n}} \cdot\mathbf{r}^{\prime}$ in the
phase, with $\mathbf{\hat{n}}=\mathbf{r} /r $ being the direction of
observation. Note that in the Green function phase
\begin{equation}
n(\lambda{z})\omega\left\vert \mathbf{r}-\mathbf{r}^{\prime}\right\vert
\simeq\omega r\left[  1-\frac{\mathbf{\hat n}\cdot\mathbf{r}^{\prime}}%
{r}+\lambda\tilde{\xi}\omega-\lambda\tilde{\xi}\omega\frac{\mathbf{\hat
n}\cdot\mathbf{r}^{\prime}}{r}+\frac{1}{2}\left(  \frac{r^{\prime}}{r}\right)
^{2}\right]  , \label{PHASE}%
\end{equation}
we are taking the radiation approximation, which means that the subdominant
terms of order $(r^{\prime}/r)^{2}$ or higher are neglected. Consistency
demands that terms proportional to the LIV parameter $\tilde{\xi}$ are larger
than the neglected one in order to properly include them in this phase. Our
general results are presented in this full far-field approximation.

Using (\ref{GRETGF}) we obtain
\begin{equation}
\mathbf{A}^{\lambda}(\omega,\mathbf{r})=4\pi\int d^{3}\mathbf{r}^{\prime
}\;G_{ret}^{\lambda}\mathbf{(}\omega\mathbf{,r-r}^{\prime}\mathbf{)\;j}%
_{T}^{\lambda}\mathbf{(}\omega\mathbf{,\mathbf{r}^{\prime})=}\frac{1}{R}%
\frac{n(\lambda z)}{\sqrt{1+z^{2}}}e^{in(\lambda z)\omega r}\int
d^{3}\mathbf{r}^{\prime}\;e^{-i\left[  n(\lambda z)\omega\mathbf{\hat{n}%
}\right]  \cdot\mathbf{r}^{\prime}}\mathbf{\;j}_{T}^{\lambda}\mathbf{(}%
\omega\mathbf{,\mathbf{r}^{\prime})},
\end{equation}
and thus we finally get
\begin{equation}
\mathbf{A}^{\lambda}(\omega,\mathbf{r})=\frac{1}{r}\frac{n(\lambda{z})}%
{\sqrt{1+{z^{2}}}}e^{in(\lambda{z})\omega r}\mathbf{j}^{\lambda}%
(\omega\mathbf{{k}_{\lambda}}) \label{APOL}%
\end{equation}
in the radiation approximation. The fields $\mathbf{A}^{+}(\omega,
\mathbf{r})$ and $\mathbf{A}^{-}(\omega,\mathbf{r})$ correspond to right and
left circular polarization respectively. Let us emphasize that the momenta
\begin{equation}
\mathbf{k}_{\lambda}=n(\lambda z)\ \omega\ \mathbf{\mathbf{\hat{n} }}
\label{KLAMBDA}%
\end{equation}
in Eq. (\ref{APOL}) are fixed in terms of the frequency and the
direction of observation. As usual, the integration over
$d^{3}\mathbf{r}^{\prime}$ has been conveniently written as the
Fourier transform in momentum space $\mathbf{j}
_{T}^{\lambda}(\omega,\mathbf{k}_{\lambda})$ of the function
$\mathbf{j} _{T}^{\lambda}(\omega,\mathbf{r}^{\prime})$ in
coordinate space. In the RHS of Eq. (\ref{APOL}) we have used the
relation (\ref{JPM}) for the transverse current. The full vector
potential is given by the superposition
\begin{equation}
\mathbf{A}(\omega,\mathbf{r})=\frac{1}{r}\frac{1}{\sqrt{1+{z^{2}}}}%
\sum_{\lambda=\pm1}n(\lambda{z})e^{in(\lambda{z})\omega r}\mathbf{j}^{\lambda
}(\omega,\mathbf{{k}_{\lambda}}). \label{VECPOT}%
\end{equation}

Hence the electric and magnetic fields are
\begin{align}
\mathbf{B(}\omega,\mathbf{r})  &  =\mathbf{\nabla}\times\mathbf{A(}%
\omega,\mathbf{r})\nonumber\\
&  =\frac{1}{r}\frac{\omega}{\sqrt{1+z^{2}}}\left[  n^{2}(z)e^{in(z)\omega
r}\mathbf{j}^{+}\mathbf{(}\omega,\mathbf{k}_{+})-n^{2}(-z)e^{in(-z)\omega
r}\mathbf{j}^{-}\mathbf{(}\omega,\mathbf{k}_{-})\right]  ,\\
\mathbf{E(}\omega,\mathbf{r})  &  =i\omega\mathbf{A(}\omega,r)\nonumber\\
&  =\frac{1}{r}\frac{i\omega}{\sqrt{1+z^{2}}}\left[  n(z)e^{in(z)\omega
r}\mathbf{j}^{+}\mathbf{(}\omega,\mathbf{k}_{+})+n(-z)e^{in(-z)\omega
r}\mathbf{j}^{-}\mathbf{(}\omega,\mathbf{k}_{-})\right]  . \label{EFIELD}%
\end{align}
Note that, contrary to the standard case, where $\mathbf{\hat{n}\times E}
\propto\mathbf{B}$, here we have
\begin{equation}
\mathbf{\hat{n}\times E}(\omega,\mathbf{r})=\frac{1}{\sqrt{1+z^{2}}}\left[
\mathbf{B}(\omega,\mathbf{r})+iz\mathbf{E}(\omega,\mathbf{r})\right]  .
\end{equation}

\section{The angular distribution of the power spectrum}

This is defined as $\frac{d^{2}P(T)}{dwd\Omega}$, where $P(T)$ is the radiated
power at time $T$ into the solid angle $d\Omega$ in a given radiation problem.
We can compute the total energy emitted in terms of the Poynting vector
(\ref{pv})%
\begin{equation}
E=\int_{-\infty}^{\infty}dt\;\mathbf{n\cdot S(}t,\mathbf{r)\;}r^{2}d\Omega.
\end{equation}
This last expression can be rewritten introducing the Fourier transform of the
Poynting vector,
\begin{equation}
E=\int_{0}^{\infty}d\omega\int d\Omega\;\frac{d^{2}E}{d\Omega d\omega}%
=\int_{0}^{\infty}\frac{d\omega}{2\pi}\left[  \mathbf{n\cdot S(}%
\omega,\mathbf{r)+n\cdot S(-}\omega,\mathbf{r)}\right]  r^{2}d\Omega,
\end{equation}
and allows us to obtain the angular distribution of the energy spectrum
\begin{equation}
\frac{d^{2}E}{d\Omega d\omega}=\frac{r^{2}}{2\pi}\left[  \mathbf{n\cdot
S(}\omega,\mathbf{r)+n\cdot S(-}\omega,\mathbf{r)}\right]  , \label{eo}%
\end{equation}
from where the angular distribution of the power spectrum can be identified
as
\begin{equation}
\frac{d^{2}E}{d\Omega d\omega}=\int_{-\infty}^{+\infty} dT\;\frac{d^{2}%
P(T)}{dwd\Omega}. \label{ep}%
\end{equation}
The Poynting vector is given by Eq. (\ref{pv}), from which results%
\begin{equation}
\mathbf{S}(\omega,\mathbf{r})=\frac{1}{4\pi}\left(  \mathbf{E(-}%
\omega,\mathbf{r)}\times\mathbf{B\mathbf{(}}\omega\mathbf{,r\mathbf{)}%
+}i\omega\tilde{\xi}\mathbf{E(-}\omega,\mathbf{r)\times E(}\omega
,\mathbf{r)}\right)  .
\end{equation}
This expression, exact in $\tilde{\xi}$, can also be written in terms of the
potentials, in which case we have
\begin{align}
\mathbf{S}(\omega,\mathbf{r})  &  \mathbf{=}\mathbf{-}\frac{i}{4\pi}\left\{
\omega^{2}\mathbf{A}(-\omega,\mathbf{r})\times\left[  n( \tilde{\xi}\omega)
\mathbf{A}_{+}(\omega,\mathbf{r})-n( -\tilde{\xi}\omega) \mathbf{A}_{-}%
(\omega,\mathbf{r})\right]  \right. \nonumber\\
&  \ \ \ \ \ \ \ \ \ \left.  -\tilde{\xi}\omega^{3}\mathbf{A}(-\omega
,\mathbf{r})\mathbf{\times A}(\omega,\mathbf{r})\right\}  .
\end{align}
Finally, using the relations (\ref{RELJPM}) of Appendix A we get
\begin{equation}
\mathbf{n\cdot S}\left(  \omega,\mathbf{r}\right)  =\frac{\omega^{2}}{4\pi
}\sqrt{1+z^{2}}\left[  \mathbf{A}_{-}(-\omega,\mathbf{r})\cdot\mathbf{A}%
_{+}(\omega,\mathbf{r})+\mathbf{A}_{+}(-\omega,\mathbf{r})\cdot\mathbf{A}%
_{-}(\omega,\mathbf{r})\right]  . \label{nso}%
\end{equation}
Introducing Eq. (\ref{nso}) in Eq. (\ref{eo}) we finally have
\begin{equation}
\frac{d^{2}E}{d\Omega d\omega}=\frac{r^{2}\omega^{2}}{4\pi^{2}}\sqrt{1+z^{2}%
}\left[  \mathbf{A}_{-}(-\omega,\mathbf{r})\cdot\mathbf{A}_{+}(\omega
,\mathbf{r})+\mathbf{A}_{-}(\omega,\mathbf{r})\cdot\mathbf{A}_{+}%
(-\omega,\mathbf{r})\right]  .
\end{equation}
Now we need to express the products $\mathbf{A}_{\mp}(-\omega,\mathbf{r}
)\cdot\mathbf{A}_{\pm}(\omega,\mathbf{r})$ in terms of the current
$\mathbf{j}(\omega,\;\mathbf{k})$ via the relation (\ref{APOL}). To this end
it is convenient to introduce the projectors (\ref{PROY}) that satisfy the
relations (\ref{AP1}), (\ref{AP2}) and (\ref{AP3}). Using these results
together with the general relation
\begin{equation}
j_{k}(-\omega,-\mathbf{k})=j_{k}^{\ast}(\omega,\mathbf{k}),
\end{equation}
we obtain
\begin{equation}
\mathbf{A}_{-}(\mp\omega,\mathbf{r})\cdot\mathbf{A}_{+}(\pm\omega
,\mathbf{r})=\frac{1}{r^{2}}\frac{n^{2}(z)}{1+z^{2}}\ j_{l}^{\ast}\left(
\omega,\omega n(\pm z)\mathbf{\hat{n}})\right)  \;P_{lk}^{\pm}\;j_{k}%
(\omega,\omega n(\pm z)\mathbf{\hat{n}}),
\end{equation}
which leads to
\begin{align}
\frac{d^{2}E}{d\Omega d\omega}  &  =\frac{1}{4\pi^{2}}\frac{\omega^{2}%
}{1+z^{2}} \left[  \frac{}{} n^{2}(z)\ j_{l}^{\ast}\left(  \omega,{}%
\mathbf{k}_{+}\right)  \;P_{lk}^{+}\;j_{k}(\omega,\mathbf{k}_{+})
+n^{2}(-z)\;j_{k}^{\ast }(\omega,
\mathbf{k}_{-})\;P_{kl}^{-}\;j_{l}\left(
\omega,\mathbf{k}_{-}\right) \frac{}{}
\right]  . \label{es}%
\end{align}

In order to identify the angular distribution of the power spectrum we need to
go back to Eq. (\ref{ep}). Each contribution in Eq. (\ref{es}) is of the type
\begin{equation}
C\left(  \omega\right)  =j_{k}^{\ast}\left(  \omega,\mathbf{k}\right)
\ X_{kr}\;j_{r}\left(  \omega,\mathbf{k}\right)  =\int_{-\infty}^{+\infty}
dt\ dt^{\prime}e^{-i\omega\left(  t-t^{\prime}\right)  }j_{k}^{\ast}\left(
t,\mathbf{k}\right)  \ X_{kr}\;j_{r}\left(  t^{\prime},\mathbf{k}\right)  .
\end{equation}
Introducing the new time variables $\tau=t-t^{\prime}$ and $T=\left(
t+t^{\prime}\right)  /2$ we get%
\begin{equation}
C\left(  \omega\right)  =\int_{-\infty}^{+\infty} dT\int_{-\infty}^{\infty
}d\tau e^{-i\omega\tau}j_{k}^{\ast}\left(  T+\tau/2,\mathbf{k}\right)
\ X_{kr}\;j_{r}\left(  T-\tau/2,\mathbf{k}\right)  .
\end{equation}
Inserting this last relation in Eq. (\ref{es}) and comparing with Eq.
(\ref{ep}) we obtain the final expression for the angular distribution of the
radiated power spectrum
\begin{align}
\frac{d^{2}P(T)}{d\omega d\Omega}&=\frac{1}{4\pi^{2}}\frac{\omega^{2}%
}{\sqrt{1+z^{2}}}\int_{-\infty}^{\infty}d\tau\ e^{-i\omega\tau} \left[
\frac{}{} n^{2}(z)j_{k}^{\ast} \left(T+\tau/2,\mathbf{k}_{+} \right)  \ P_{kr}%
^{+}\;j_{r}\left(  T-\tau/2,\mathbf{k}_{+} \right)  \right. \nonumber\\
&  \left.  +n^{2}(-z)j_{k}^{\ast}\left(  T+\tau/2,\mathbf{k}_{-} \right)  \ P_{kr}%
^{-}\;j_{r}\left(  T-\tau/2,\mathbf{k}_{-} \right)  \frac{}{} \right]  , \label{poo}%
\end{align}
as the sum of the contributions of both circular polarizations.

\section{Synchrotron radiation}

Let us rewrite Eq. (\ref{poo}) in the form%
\begin{align}
\frac{d^{2}P(T)}{d\omega d\Omega}  &  =\sum_{\lambda=\pm1}\frac{\omega^{2}%
}{4\pi^{2}}\frac{n^{2}(\lambda z)}{\sqrt{1+z^{2}}}
\int_{-\infty}^{\infty }d\tau e^{-i\omega\tau}j_{k}^{\ast}\left(
T+\tau/2,\mathbf{k}_{\lambda}\right) \
P_{kr}^{\lambda}\;j_{r}\left(T-\tau/2,\mathbf{k}_{\lambda}\right)
,
\end{align}
where the current associated to the motion of a particle in an external
magnetic field was already obtained in Eq. (\ref{CURRENT}). Its spatial
Fourier transform is
\begin{equation}
\mathbf{j}(t,\mathbf{k})=\int d^{3}y{} (t,\mathbf{r})e^{-i\mathbf{k}\cdot\mathbf{y}%
}=q\mathbf{v(}t\mathbf{)}e^{-i\mathbf{k}\cdot\mathbf{r(}t\mathbf{)}},
\label{SPFT}%
\end{equation}
and thus we get%
\begin{equation}
\frac{d^{2}P(T)}{d\omega d\Omega}=\sum_{\lambda=\pm1}\frac{\omega^{2}}%
{2\pi^{2}}\frac{n^{2}(\lambda z)}{\sqrt{1+z^{2}}}\,\mathcal{J}(T,\omega
,\lambda,\mathbf{\hat{n}}), \label{pe}%
\end{equation}
where
\begin{equation}
\mathcal{J}(T,\omega,\lambda,\mathbf{n})=\int_{-\infty}^{\infty}d\tau
e^{-i\omega\tau}j_{k}^{\ast}\left(T+\tau/2,\mathbf{k}_{\lambda}\right)
\
P_{kr}^{\lambda}\;j_{r}\left(T-\tau/2,\mathbf{k}_{\lambda}\right)
.
\end{equation}
This integral can be expressed as
\begin{equation}
\mathcal{J}(T,\omega,\lambda,\mathbf{n})=q^{2}\int_{-\infty}^{\infty}d\tau
e^{-i\omega\tau}e^{i\omega n(\lambda z)\mathbf{\hat{n}}\cdot\left(
\mathbf{r(}T+\tau/2\mathbf{)-r(}T-\tau/2\mathbf{)}\right)  }\mathfrak{J},
\label{SJOTA}%
\end{equation}
with%
\begin{align}
\mathfrak{J}  &  =v_{k}(T+\tau/2)\ P_{kr}^{\lambda}\;v_{r}(T-\tau
/2)\nonumber\\
&  =\frac{1}{2}\left[  \mathbf{v}(T+\tau/2)\cdot\mathbf{v}(T-\tau
/2)-\mathbf{\hat{n}\cdot v}(T+\tau/2)\;\mathbf{\hat{n}}\cdot\mathbf{v}%
(T-\tau/2) -i\lambda\mathbf{\hat{n}\cdot v}(T+\tau/2)\times\mathbf{v}%
(T-\tau/2)\right]  . \label{v}%
\end{align}
Here it is useful to rewrite the terms containing $\mathbf{\hat{n}\cdot j\;}
\approx\mathbf{\hat{n}\cdot v}$ in terms of the charge density. To this end we
use current conservation in momentum space
\begin{equation}
\omega\rho(\omega,\mathbf{k})-\mathbf{k\cdot j}(\omega,\mathbf{k})=0,
\end{equation}
obtaining
\begin{equation}
\mathbf{\hat{n}\cdot
j}(\omega,\mathbf{k}_{\lambda})=\frac{1}{n(\lambda z)}\rho
(\omega,\mathbf{k}_{\lambda}),
\end{equation}
which leads to
\begin{equation}
\mathbf{\hat{n}\cdot j}(t,\mathbf{k}_{\lambda})=\frac{1}{n(\lambda
z)}\rho (t,\mathbf{k}_{\lambda}).
\end{equation}
In this way we have
\begin{align}
\left[  \mathbf{\hat{n}}\cdot\mathbf{j}^{\ast}\left(
T+\tau/2,\mathbf{k}_{\lambda }\right)  \right]  \, \left[
\mathbf{\hat{n}}\cdot\mathbf{j}\left(T-\tau/2,\mathbf{k}_{\lambda}\right)
\right]=\frac{1}{n^{2}(\lambda z)}\rho^{\ast }\left[ T+\tau/2,
\mathbf{k}_{\lambda}\right]  \rho\left[T-\tau/2,
\mathbf{k}_{\lambda }\right] .
\end{align}
In other words, in the expression of $\mathfrak{J}$, Eq. (\ref{v}), we can
make the replacement $\mathbf{\hat{n}\cdot v(t)\, \rightarrow} \,1/n(\lambda
z), $ leading to
\begin{equation}
\mathfrak{J}=\frac{1}{2}\left[  \mathbf{v}\left(  T+\tau/2\right)
\cdot\mathbf{v}\left(  T-\tau/2\right)  -n^{-2}(\lambda z)-i\lambda
\mathbf{\hat{n} \cdot v}\left(  T+\tau/2\right)  \times\mathbf{v}\left(
T-\tau/2\right)  \right]  . \label{v1}%
\end{equation}
In the following we will consider the simplest case, when the orbit of the
charged particle is in a plane perpendicular to the magnetic field, i.e. the
case where $\beta_{\Vert}=0$ and $\beta_{\bot}=\beta$. To continue with the
calculation we introduce the direction of observation
\begin{equation}
\mathbf{\hat{n}}=(\sin\theta,0,\cos\theta), \label{cs}%
\end{equation}
in the same coordinate system where $\mathbf{r}(t)$ and $\mathbf{v}(t)$ were
defined in Eqs. (\ref{rt}) and (\ref{vt}). Recalling the corresponding
expressions we calculate
\begin{align}
\mathbf{v}(T+\tau/2)\cdot\mathbf{v}(T-\tau/2)  &  =\beta^{2}\cos\omega_{0}%
\tau,\\
\mathbf{n\cdot v}(T+\tau/2)\times\mathbf{v}(T-\tau/2)  &  =-\beta^{2}%
\cos\theta\sin\omega_{0}\tau,
\end{align}
and substituting in Eq. (\ref{v1}) we are left with%
\begin{equation}
\mathfrak{J}=\frac{1}{2}\left[  \beta^{2}\cos\omega_{0}\tau-n^{-2}(\lambda
z)+i\lambda\beta^{2}\cos\theta\sin\omega_{0}\tau\right]  . \label{scriptj}%
\end{equation}

Now we can compute the cycle average over the macroscopic time $T$, of the
angular distribution of the radiated power spectrum
\begin{equation}
\left\langle \frac{d^{2}P(T)}{d\omega d\Omega}\right\rangle =\sum_{\lambda
=\pm1}\frac{\omega^{2}}{4\pi^{2}}\frac{n^{2}(\lambda z)}{\sqrt{1+z^{2}}%
}\left\langle \mathcal{J}(T,\omega,\lambda,\mathbf{\hat{n}})\right\rangle .
\label{ps1}%
\end{equation}
The dependence on $T$ is in the exponential factor of Eq. (\ref{SJOTA}), where
we have
\begin{equation}
\mathbf{\hat{n}}\cdot\left[  \mathbf{r(}T+\tau/2\mathbf{)-r(}T-\tau
/2\mathbf{)}\right]  =-2R\sin\theta\sin\left(  \omega_{0}T\right)  \sin\left(
\omega_{0}\tau/2\right)  .
\end{equation}
Thus the required averaged value is
\begin{equation}
<e^{i\omega n(\lambda z)\mathbf{\hat{n}}\cdot\left(  \mathbf{r(}%
T+\tau/2\mathbf{)-r(}T-\tau/2\mathbf{)}\right)  }>=J_{0}(2Y\sin\theta),
\end{equation}
with
\begin{equation}
Y=\omega\ n(\lambda z)\ R\;\sin\left(  \omega_{0}\tau/2\right)  . \label{YE}%
\end{equation}
From here we get%
\begin{align}
<\mathcal{J}(T,\omega,\lambda,\mathbf{n})>  &  =\frac{q^{2}}{2}\int_{-\infty
}^{\infty}d\tau\;e^{-i\omega\tau}J_{0}(2Y\sin\theta)\nonumber\\
&  \times\left[  \beta^{2}\cos\omega_{0}\tau-n^{-2}(\lambda z)+i\lambda
\beta^{2}\cos\theta\;\sin\left(  \omega_{0}\tau\right)  \right]  . \label{AVJ}%
\end{align}
Now we apply the addition theorem for Bessel functions \cite{ADTHEO}
\begin{equation}
J_{0}\left(  k\sqrt{\rho^{2}+\sigma^{2}-2\rho\sigma\cos(\phi-\phi^{\prime}%
)}\right)  =\sum_{m=-\infty}^{\infty}J_{m}(k\rho)J_{m}(k\sigma)e^{im(\phi
-\phi^{\prime})},
\end{equation}
with%
\begin{equation}
\rho=\sigma=R,\;\;\ k=\omega n(\lambda z)\sin\theta,\;\;\;(\phi-\phi^{\prime
})=\omega_{0}\tau,
\end{equation}
which yields
\begin{equation}
J_{0}\left(  2Y\sin\theta\right)  =\sum_{m=-\infty}^{\infty}J_{m}^{2}\left[
\omega n(\lambda z)R\sin\theta\right]  e^{im\omega_{0}\tau}.
\end{equation}
The next step is to rewrite the trigonometric functions depending upon
$\omega_{0}\tau$ in Eq. (\ref{scriptj}) as exponentials and to perform the
$\tau$ integration. In such a way we finally arrive at the expression for the
observed time average of the power angular distribution, given by
\begin{equation}
\left\langle \frac{d^{2}P(T)}{d\omega d\Omega}\right\rangle =\sum_{\lambda
=\pm1}\sum_{m=0}^{\infty}\delta({\omega}-m\omega_{0})\frac{dP_{m,\lambda}%
}{d\Omega},
\end{equation}
with
\begin{align}
\frac{dP_{m,\lambda}}{d\Omega}  &  =\frac{\omega_{m}^{2}q^{2}}{8\pi}%
\frac{n^{2}(\lambda z _{m})}{\sqrt{1+\left(  z_{m}\right)  ^{2}}}\left\{
\frac{}{} \beta^{2}\left[  \left(  1+\lambda\cos\theta\right)  J_{m-1}%
^{2}(W_{m\lambda})+\left(  1-\lambda\cos\theta\right)  J_{m+1}^{2}%
(W_{m\lambda})\right]  \right. \nonumber\\
&  \left.  -2[n(\lambda z_{m})]^{-2} J_{m}^{2}(W_{m\lambda})\frac{}{}
\right\}  , \label{ADMH}%
\end{align}
where $\omega_{m}=m\omega_{0}, \, z_{m}= {\tilde\xi} \omega_{m}$ and
\begin{equation}
W_{m\lambda}=\omega_{m}n(\lambda z_{m})R\sin\theta. \label{WLM}%
\end{equation}
Using the relation $\omega_{0}R=\beta$ we can rewrite $W_{m\lambda}$ as
\begin{equation}
W_{m\lambda}=m\,n(\lambda z_{m})\beta\sin\theta.
\end{equation}
We notice that in the limit $\xi=0$ the unpolarized angular distribution of
the power obtained from Eq. (\ref{ADMH}) coincides with the standard result
given in Eq. (38.37) of Ref. \cite{SCHWBOOK}. Let us observe that to first
order in the LIV parameters we can take
\begin{equation}
\tilde{\xi}\omega_{0}=\tilde{\xi}\frac{|q\mathbf{B}|}{\mu\gamma},
\end{equation}
according to Eqs. (\ref{OMEGA0}) and (\ref{18}), since $\tilde{\xi}\tilde
{\eta}$ is of second order in the correction parameters.

Using additional properties of the Bessel functions and following Ref.
\cite{SCHWBOOK} we can recast Eq. (\ref{ADMH}) as
\begin{align}
\frac{dP_{m\lambda}}{d\Omega}  &  =\frac{\omega_{m}^{2}q^{2}}{4\pi}\frac
{n^{2}(\lambda z_{m})}{\sqrt{1+z_{m}^{2}}}\left\{  \left[  \beta J_{m}%
^{\prime}(W_{m\lambda})\right]  ^{2}+\left[  n^{-1}(\lambda z_{m}%
)\,cot\theta\,J_{m}(W_{m\lambda})\right]  ^{2}\right\} \nonumber\\
&  +\lambda\frac{\omega_{m}^{2}q^{2}\beta}{4\pi}\frac{n(\lambda z_{m})}%
{\sqrt{1+z_{m}{}^{2}}}\cot\theta\frac{dJ_{m}^{2}(W_{m\lambda})}{dW_{m\lambda}%
}. \label{pmlt}%
\end{align}
It is interesting to remark that the above expression can be written as a
perfect square
\begin{equation}
\frac{dP_{m\lambda}}{d\Omega}=\frac{\omega_{m}^{2}q^{2}}{4\pi}\frac{1}%
{\sqrt{1+z_{m}^{2}}}\left[  \lambda\beta n(\lambda z_{m})J_{m}^{\prime
}(W_{m\lambda})+ \cot\theta\,J_{m}(W_{m\lambda})\right]  ^{2}. \label{pmlt1}%
\end{equation}

\subsection{Total spectral distribution}

To calculate the integrated power spectrum it is convenient to start from Eqs.
(\ref{ps1}) and (\ref{AVJ}). Integrating over the solid angle $\Omega$ we
have
\begin{equation}
\left\langle \frac{dP(T)}{d\omega}\right\rangle =2\pi\sum_{\lambda=\pm1}%
\frac{\omega^{2}}{4\pi^{2}}\frac{n^{2}(\lambda z)}{\sqrt{1+z^{2}}}\int
\sin\theta d\theta<\mathcal{J}(T,\omega,\lambda,\mathbf{n})>, \label{ps3}%
\end{equation}
where the required integral is
\begin{align}
\int\sin\theta d\theta<\mathcal{J}(T,\omega,\lambda,\mathbf{n})>  &
=\frac{q^{2}}{2}\int_{-\infty}^{\infty}d\tau\;e^{-i\omega\tau}\nonumber\\
&  \times\left\{  \left[  \beta^{2}\cos\omega_{0}\tau-n^{-2}(\lambda
z)\right]  I_{1}+\lambda\beta^{2}\left(  i\sin\omega_{0}\tau\right)
I_{2}\right\}  .
\end{align}
Here%
\begin{align}
I_{1}  &  =\int_{0}^{\pi}d\theta\;\sin\theta\;J_{0}(2Y\sin\theta)=\frac
{\sin2Y}{Y},\\
I_{2}  &  =\int_{0}^{\pi}d\theta\;\sin\theta\;\cos\theta\;J_{0}(2Y\sin
\theta)=0,
\end{align}
where $Y$ is given in Eq. (\ref{YE}). The last integral, related to parity
violation, is zero by symmetry. Substituting in Eq. (\ref{ps3}) we get
\begin{equation}
\left\langle \frac{dP(T)}{d\omega}\right\rangle =\frac{q^{2}}{4\pi R}%
\sum_{\lambda=\pm1}\frac{\omega n(\lambda z)}{\sqrt{1+z^{2}}}\int_{-\infty
}^{\infty}d\tau e^{-i\omega\tau}f_{\lambda}(\omega_{0}\tau), \label{rp}%
\end{equation}
where%
\begin{equation}
f_{\lambda}(\phi)=\frac{\sin\left[  2\omega Rn(\lambda z)\sin\left(
\phi/2\right)  \right]  }{\sin\left(  \phi/2\right)  }\left(  \beta{\,}%
^{2}\cos\phi-n^{-2}(\lambda z)\right)  .
\end{equation}
Following the method of Ref. \cite{SCHWBOOK} we introduce the Fourier
decomposition of the periodic function $f_{\lambda}(\phi)$
\begin{equation}
f_{\lambda}(\phi)=\sum_{m=-\infty}^{\infty}e^{im\phi}f_{\lambda m},\qquad
f_{\lambda m}=\int_{-\pi}^{\pi}\frac{d\phi}{2\pi}e^{-im\phi}f_{\lambda}(\phi),
\end{equation}
and performing the $\tau$ integration we have
\begin{equation}
\left\langle \frac{dP(T)}{d\omega}\right\rangle =\frac{q^{2}\omega_{0}}%
{2R}\sum_{\lambda=\pm1}\sum_{m=0}^{\infty}\delta(\omega-m\omega_{0}%
)\frac{mn(\lambda z)}{\sqrt{1+z^{2}}}f_{\lambda m}.
\end{equation}

It only remains to compute $f_{\lambda m}$
\begin{equation}
f_{\lambda m}=\int_{-\pi}^{\pi}\frac{d\phi}{2\pi}e^{-im\phi}\frac{\sin\left[
2\omega n(\lambda z)R\;\sin\left(  \phi/2\right)  \right]  }{\sin\left(
\phi/2\right)  }\left[  \beta{\,}^{2}\cos\phi-n^{-2}(\lambda z)\right]  .
\end{equation}
The only contribution from the exponential comes from the symmetric part under
$\phi\rightarrow-\phi$. After rewriting $\cos\phi$ in terms of the $\phi/2$ we
are left with
\begin{equation}
f_{\lambda m}=\int_{0}^{\pi}\frac{d\phi}{\pi}\cos m\phi\frac{\sin\left[
2\omega n(\lambda z)R\;\sin\left(  \phi/2\right)  \right]  }{\sin\left(
\phi/2\right)  }\left[  \beta^{2}-n^{-2}(\lambda z)-2\beta^{2}\sin^{2}\left(
\phi/2\right)  \right]  .
\end{equation}

Using the following representations of the Bessel functions \cite{SCHWBOOK}
\begin{align}
J_{2m}(x)  &  =\int_{0}^{\pi}\frac{d\phi}{\pi}\cos m\phi\cos\left(
x\sin\left(  \phi/2\right)  \right)  ,\label{j2m}\\
J_{2m}^{\prime}(x)  &  =-\int_{0}^{\pi}\frac{d\phi}{\pi}\sin\left(
\phi/2\right)  \cos m\phi\;\sin\left(  x\sin\left(  \phi/2\right)  \right)
,\\
\int_{0}^{x}dyJ_{2m}(y)  &  =\int_{0}^{\pi}\frac{d\phi}{\pi}\cos m\phi
\frac{\sin\left(  x\sin\left(  \phi/2\right)  \right)  }{\sin\left(
\phi/2\right)  },
\end{align}
with $x=2R\omega n(\lambda z)=2mR\omega_{0}n(\lambda z)$, and recalling that
$z=\tilde{\xi}\omega=\tilde{\xi}m\omega_{0}$ and $R\omega_{0}=\beta$, we
arrive at
\begin{equation}
f_{\lambda m}=2\beta^{2}J_{2m}^{\prime}(2m\beta\;n(\lambda z))-\left[
n^{-2}(\lambda z)-\beta^{2}\right]  \int_{0}^{2mn(\lambda z)\beta}%
dx\;J_{2m}(x).
\end{equation}
Thus the total averaged and integrated power radiated in the $m^{th}$ harmonic
is
\begin{align}
P_{m}  &  =\frac{q^{2}m\omega_{0}}{2R\sqrt{1+z^{2}}}\left\{
n(z)\left[ 2\beta^{2}J_{2m}^{\prime}(2m\beta n(z))-\left[
n^{-2}(z)-\beta^{2}\right]
\int_{0}^{2mn(z)\beta}dx\;J_{2m}(x)\right]  \right. \nonumber\\
&  \ \ \ \ \ \ \ \ \ \ \ \ \ \ \ \ \ \ \ \left.  +n(-z)\left[  2\beta
^{2}J_{2m}^{\prime}(2m\beta n(-z))-\left[  n^{-2}(-z)-\beta^{2}\right]
\int_{0}^{2mn(-z)\beta}dx\;J_{2m}(x)\right]  \right\}  . \label{PMDO}%
\end{align}
This result is exact in $z=\tilde{\xi}\omega$. Let us observe that each
separate contribution in Eq. (\ref{PMDO}) has the form given in Eq. (2.22b) of
Ref. \cite{SCWANNP}, after the appropriate change $e^{2}/4\pi\rightarrow
q^{2}$ is made, multiplied by the additional factor
\begin{equation}
\frac{1}{2}\frac{n(\lambda z)}{\sqrt{1+z^{2}}}=\frac{n(\lambda z)}%
{n(z)+n(-z)}=\frac{n^{2}(\lambda z)}{1+n^{2}(\lambda z)}, \label{ADDFACT}%
\end{equation}
respectively. The calculation performed in this reference is for a constant
index of refraction but, as mentioned in footnote 3, the corresponding results
hold for an $\omega$ and $H$ dependence of $n$ and $\mu$. At this stage, the
additional factors (\ref{ADDFACT}) together with the specific form of the
refraction indices are the remainders that we are considering a different electrodynamics.

\subsection{Polarization of the synchrotron radiation}

In general, the electric field (\ref{EFIELD}), which is naturally written in
terms of the circular polarization basis\ $\mathbf{e}_{\pm}$ introduced in
Appendix B, will exhibit elliptic polarization. To describe the polarization
we will use the Stokes parameters according to the definitions of Ref.
\cite{RL}
\begin{align}
I  &  =|\left(  \mathbf{e}_{+}^{\ast}\cdot\mathbf{E(}\omega\mathbf{,r)}%
\right)  |^{2}+|\left(  \mathbf{e}_{-}^{\ast}\cdot\mathbf{E(}\omega
\mathbf{,r)}\right)  |^{2},\\
V  &  =|\left(  \mathbf{e}_{+}^{\ast}\cdot\mathbf{E(}\omega\mathbf{,r)}%
\right)  |^{2}-|\left(  \mathbf{e}_{-}^{\ast}\cdot\mathbf{E(}\omega
\mathbf{,r)}\right)  |^{2},\\
Q  &  =2{Re}\left[  \left(  \mathbf{e}_{+}^{\ast}\cdot\mathbf{E(}%
\omega\mathbf{,r)}\right)  ^{\ast}\left(  \mathbf{e}_{-}^{\ast}\cdot
\mathbf{E(}\omega\mathbf{,r)}\right)  \right]  ,\\
U  &  =2{Im}\left[  \left(  \mathbf{e}_{+}^{\ast}\cdot\mathbf{E(}\omega
_{m}\mathbf{,r)}\right)  ^{\ast}\left(  \mathbf{e}_{-}^{\ast}\cdot
\mathbf{E(}\omega_{m}\mathbf{,r)}\right)  \right]  . \label{STOKES}%
\end{align}
It is convenient to define the reduced Stokes parameters by dividing each one
by the intensity $I$, in such a way that
\begin{equation}
v=\frac{V}{I},\;\;q=\frac{Q}{I},\;\;u=\frac{U}{I},
\end{equation}
satisfying the relation
\begin{equation}
1=v^{2}+q^{2}+u^{2}. \label{POLESQ}%
\end{equation}
As usual, purely circular polarization is described by $v=1$, $q=u=0$, while
purely linear polarization corresponds to $v=0$.

A first step in the description of the polarization is the
explicit calculation of the electric field, which is given in
terms of the currents $\mathbf{j}^{+}(\omega, \mathbf{k}_{+})$ and
$\mathbf{j}^{-}(\omega, \mathbf{k}_{-})$. We need to calculate the
time Fourier transform of the current (\ref{SPFT}) for each
polarization $\lambda$ and subsequently project into the circular
basis. We start from
\begin{equation}
\mathbf{j\,(}t\mathbf{,k}_{\lambda}\mathbf{)}\mathbf{=}q \mathbf{v(}%
t\mathbf{)}e^{-i\mathbf{k}_{\lambda}\mathbf{\cdot r(}t\mathbf{)}}%
\mathbf{=}q\beta \left(
-\sin\omega_{0}t,\,\cos\omega_{0}t,\,0\right)
e^{-i\mathbf{k}_{\lambda}\mathbf{\cdot r(}t\mathbf{)}},
\end{equation}
where we can write
\begin{equation}
e^{-i\mathbf{k}_{\lambda}\mathbf{\cdot r(}t\mathbf{)}}=e^{-i\frac{\omega
}{\omega_{0}}n\mathbf{(}\lambda z\mathbf{)}\beta\sin\theta\cos\omega_{0}t}.
\end{equation}
The Fourier transform is
\begin{align}
\mathbf{j(}\omega,\mathbf{k}_{\lambda}\mathbf{)}  &  \mathbf{=}\int_{-\infty
}^{\infty}dt\ e^{i\omega t}\;\mathbf{j(}t,\mathbf{k}_{\lambda})\nonumber\\
&  =q\beta \int_{-\infty}^{\infty}dt\ e^{i\omega t}\left(
-\sin\omega _{0}t,\cos\omega_{0}t,0\right)
e^{-iW_{\lambda}\cos\omega_{0}t},
\end{align}
with $W_{\lambda}=\left(  \omega\ n(\lambda z)\beta\sin\theta\right)
/\omega_{0}$. Using the generating function for the Bessel functions of
integer order (see for example Eq. (38.27) in Ref. \cite{SCHWBOOK}) we obtain
\begin{equation}
e^{-iW_{\lambda}\cos\omega_{0}t}=\sum_{m=-\infty}^{\infty}(-i)^{m}%
e^{-im\omega_{0}t}J_{m}(W_{\lambda}),
\end{equation}
and in this way we have
\begin{equation}
\mathbf{j(}\omega\mathbf{,k}_{\lambda}\mathbf{)}\mathbf{=}q\beta
\sum_{m=-\infty}^{\infty}\;(-i)^{m}J_{m}(W_{\lambda})\int_{-\infty}^{\infty
}e^{it\left(  \omega-m\omega_{0}\right)  }\left(  -\sin\omega_{0}%
t,\;\cos\omega_{0}t,\;0\right)  \;dt.
\end{equation}
Now we can perform the integration with respect to $t$. The final result can
be written in terms of each mode contributing to the current
\begin{equation}
\mathbf{j(}\omega\mathbf{,k}_{\lambda}\mathbf{)}\mathbf{=}\sum_{m=-\infty
}^{\infty}\delta(\omega-m\omega_{0})\,\mathbf{j}_{m}(\omega_{m}\mathbf{,k}%
_{\lambda}),
\end{equation}
where
\begin{equation}
\mathbf{j}_{m}(\omega_{m},{\mathbf{k}}_{\lambda})=2\pi(-i)^{m}q\,\beta
\left[  \frac{m}{W_{\lambda
m}}J_{m}(W_{m\lambda}),\;i\frac{d}{dW_{\lambda
m}}J_{m}(W_{m\lambda}),0\right]  ,
\end{equation}
with $W_{m\lambda}$ given by Eq. (\ref{WLM}).

The projection $\mathbf{j}_{m}^{\pm}(\omega_{m},\mathbf{k}_{\lambda})$ of the
currents into the circular basis is given by Eqs. (\ref{PROYC}) of the
Appendix. Using these relations together with the definitions (\ref{EFIELD})
we arrive at the final expressions for the circularly polarized electric
fields%
\begin{align}
\mathbf{E}_{m}^{+}(\omega_{m},\mathbf{k}_{+})  &  =-\sqrt{2}\pi(-i)^{m}%
\frac{q\beta}{r}\frac{\omega_{m} n(z_{m})}{\sqrt{1+z_{m}^{2}}}e^{in(z_{m}%
)\omega_{m} r}\left(  B_{m}^{+}+A_{m}^{+}\cos\theta\;\right)  \mathbf{e}%
_{+},\\
\mathbf{E}_{m}^{-}(\omega_{m}, \mathbf{k}_{-})  &  =-\sqrt{2}\pi(-i)^{m}%
\frac{q\beta}{r}\frac{\omega_{m} n(-z_{m})}{\sqrt{1+z_{m}^{2}}}e^{in(-z_{m}%
)\omega_{m} r}\left(  B_{m}^{-}-A_{m}^{-}\cos\theta\;\right)  \mathbf{e}_{-},
\label{CPOLEF}%
\end{align}
where the notation is
\begin{equation}
\mathbf{j}_{m}(\omega_{m},\mathbf{k}_{\lambda}\mathbf{)=}2\pi(-i)^{m}q
\left[ A_{m}^{\lambda},\;iB_{m}^{\lambda},\;0\right]  ,
\end{equation}
with
\begin{equation}
A_{m}^{\lambda}=\beta\frac{m}{W_{\lambda m}}J_{m}(W_{m\lambda}),\;\;B_{m}%
^{\lambda}=\beta\frac{dJ_{m}(W_{m\lambda})}{dW_{m\lambda}}.
\end{equation}
The combination $A^{\lambda}\cos\theta$ reduces to%
\begin{equation}
A_{m}^{\lambda}\cos\theta=D_{m}^{\lambda}=\frac{J_{m}(W_{m\lambda})}%
{\tan\theta}.
\end{equation}
In the standard situation, where ${\tilde\xi}=0$, the above expressions
produce the result
\begin{align}
\mathbf{E}_{m}(\omega_{m},\,\omega_{m}\,\mathbf{n)}  &  =2\pi(-i)^{m}%
q\,\omega_{m}\frac{e^{i\omega_{m}r}}{r}\nonumber\\
&  \times\left[  \frac{\cos^{2}\theta}{\sin\theta}\,J_{m}(W_{m\lambda
}),\;i\beta\frac{dJ_{m}(W_{m\lambda})}{dW_{m\lambda}},\;-\cos\theta
J_{m}(W_{m\lambda})\right]  ,
\end{align}
which is clearly orthogonal to $\mathbf{\hat{n}}$ and parallel to the
corresponding electric field given in Eq. (6.65) of Ref. \cite{PERATT} for the
case $\beta_{\parallel}=0$.

Next we calculate the Stokes parameters, Eqs. (\ref{STOKES}), starting from
Eqs. (\ref{CPOLEF}). The required projections are
\begin{align}
\left(  \mathbf{e}_{+}^{\ast}\cdot\mathbf{E}_{m}(\omega_{m},\mathbf{r)}%
\right)   &  =\mathbf{-}\sqrt{2}\pi(-i)^{m}\frac{q}{r}\frac{\omega_{m}
n(z_{m})}{\sqrt{1+z_{m}^{2}}}e^{in(z_{m})\omega r}\left(  B_{m}^{+}+D_{m}%
^{+}\right)  ,\\
\left(  \mathbf{e}_{-}^{\ast}\cdot\mathbf{E}_{m}(\omega_{m},\mathbf{r)}%
\right)   &  =\mathbf{-}\sqrt{2}\pi(-i)^{m}\frac{q}{r}\frac{\omega_{m}
n(-z_{m})}{\sqrt{1+z_{m}^{2}}}e^{in(-z_{m})\omega r}\left(  B_{m}^{-}%
-D_{m}^{-}\right)  ,
\end{align}
and we also need
\begin{align}
\left(  \mathbf{e}_{+}^{\ast}\cdot\mathbf{E}_{m}\mathbf{(}\omega
_{m}\mathbf{,r)}\right)  ^{\ast}\left(  \mathbf{e}_{-}^{\ast}\cdot
\mathbf{E}_{m}\mathbf{(}\omega_{m}\mathbf{,r)}\right)   &  =2\left(  \frac{\pi
q\,\omega_{m} }{r}\right)  ^{2}\frac{n(z_{m})n(-z_{m})}{1+z_{m}^{2}%
}e^{i\left[  n(-z_{m})-n(z_{m})\right]  \omega r}\nonumber\\
&  \times\left(  B_{m}^{+}+D_{m}^{+}\right)  \left(  B_{m}^{-}-D_{m}%
^{-}\right)  .
\end{align}
In this way we arrive at the reduced combinations
\begin{equation}
v_{m}=\frac{1-R_{m}^{2}}{1+R_{m}^{2}},\;\;\;q_{m}=\frac{2R_{m}}{1+R_{m}^{2}%
}\cos\left\{  \frac{}{} \left[  n(-z_{m})-n(z_{m})\right]  \omega_{m}
r\right\}  ,\;\;u_{m}=\frac{2R_{m}}{1+R_{m}^{2}}\sin\left\{  \frac{}{} \left[
n(-z_{m})-n(z_{m})\right]  \omega_{m} r\right\}  , \label{sp}%
\end{equation}
which completely characterize the polarization of the wave in terms of the
polarization index
\begin{equation}
R_{m}=\frac{n(-z_{m})}{n(z_{m})}\frac{\left(  B_{m}^{-}-D_{m}^{-}\right)
}{\left(  B_{m}^{+}+D_{m}^{+}\right)  }. \label{POLINDEX}%
\end{equation}
From the above relations the verification of the condition (\ref{POLESQ}) is
direct. The purely circular polarization corresponds to $R_{m}=0$, while the
purely linear polarization is given by $R_{m}^{2}=1$. Substituting the
corresponding expressions for $B_{m}^{\pm}$ and $D_{m}^{\pm}$ we obtain the
general expression for the polarization index
\begin{equation}
R_{m}=\frac{n(-z_{m})}{n(z_{m})}\frac{\left(  \beta\sin\theta J_{m}^{\prime
}(W_{m-})-\cos\theta J_{m}(W_{m-})\right)  }{\left(  \beta\sin\theta
J_{m}^{\prime}(W_{m+})+\cos\theta J_{m}(W_{m+})\right)  }. \label{PINDEXP}%
\end{equation}
In the limit $\xi=0$ we have $B_{m}^{\pm}=B_{m}$ and $D_{m}^{\pm}=D_{m}$ and
Eq. (\ref{POLINDEX}) reduces to
\begin{equation}
R_{m}=\frac{B_{m}-D_{m}}{B_{m}+D_{m}}=\frac{1-\frac{D_{m}}{B_{m}}}%
{1+\frac{D_{m}}{B_{m}}}.
\end{equation}
The additional definition
\begin{equation}
r_{m}=\frac{D_{m}}{B_{m}}=\frac{J_{m}(m\beta\sin\theta)}{\beta\tan
\theta\,J_{m}^{\prime}(m\beta\sin\theta)}%
\end{equation}
leads to
\begin{equation}
v_{m}=\frac{2r_{m}}{1+r_{m}^{2}},\qquad q_{m}=\frac{1-r_{m}^{2}}{1+r_{m}^{2}%
},\qquad u_{m}=0,
\end{equation}
which coincides, up to the sign of $r_{m}$, with the parameters introduced in
Ref. \cite{PERATT}.

\section{Dominating effects of the Lorentz violation in photon dynamics}

Several characteristics of the synchrotron radiation are affected by the
Lorentz violating parameter $\tilde{\xi}$. From the phenomenological point of
view the more relevant are the power angular distribution, the total power and
the polarization of the radiation. In what follows we discuss these effects.

\subsection{Angular distribution of the power radiated in the unpolarized
$m^{th}$-harmonic}

In this subsection we expand the distribution (\ref{pmlt}) to first order in
$\tilde{\xi}$. Notice that its general structure is given by
\begin{equation}
\frac{d\hat{P}_{m,\lambda}}{d\Omega}=F_{m}(\lambda\tilde{\xi} )+\lambda
G_{m}(\lambda\tilde{\xi} ),
\end{equation}
and thus we get%
\begin{equation}
\frac{dP_{m,\lambda}}{d\Omega}=F_{m}(0)+\tilde{\xi}G_{m}^{\prime}%
(0)+\lambda\left[  \tilde{\xi}F_{m}^{\prime}(0)+G_{m}(0)\right]  ,
\end{equation}
where the derivatives are with respect to the argument $\lambda\tilde{\xi}$ of
the functions. When we sum over polarizations it is evident that the term
linear in $\lambda$ vanishes, but nevertheless there remains a contribution
proportional to $\tilde{\xi}$. Here we identify the unpolarized average
angular distribution of the power radiated into the $m^{th}$ harmonic
\begin{equation}
\frac{dP_{m}}{d\Omega}=\frac{dP_{m+}}{d\Omega}+\frac{dP_{m-}}{d\Omega
}=2\left[  F_{m}(0)+\tilde{\xi}G_{m}^{\prime}(0)\right]  . \label{prm}%
\end{equation}
From Eq. (\ref{pmlt}), and eliminating the second derivative of the Bessel
function via its differential equation we get
\begin{align}
F_{m}(0)  &  =\frac{\left(  q \omega_{m}\right)  ^{2}}{4\pi}\left\{  \left[
\beta J_{m}^{\prime}(W_{m})\right]  ^{2}+\left[  \cot\theta\;J_{m}%
(W_{m})\right]  ^{2}\right\}, \\
G^{\prime}(0)  &  =\frac{\left(  q \omega_{m}\right)  ^{2}m \omega_{m}}{2\pi
}\cos\theta\left\{  \left[  \beta\;J_{m}^{\prime}(W_{m})\right]  ^{2}+\left(
1-\beta^{2}\sin^{2}\theta\right)  \left[  \frac{J_{m}(W_{m})}{\sin\theta
}\right]  ^{2}\right\}  ,
\end{align}
where%
\begin{equation}
W_{m}=\left.  W_{m\lambda}\right\vert _{\tilde{\xi}=0}=m\beta\sin\theta.
\end{equation}
After some rearrangement and use of trigonometric identities, we obtain from
Eq. (\ref{prm}) our final result
\begin{align}
\frac{dP_{m}}{d\Omega}  &  =\frac{\left(  q {\omega}_{m}\right)  ^{2}}{2\pi
}\left(  1+2(m \tilde{\xi}){\omega}_{m}\cos\theta\right)  \left\{  \left[
\beta J_{m}^{\prime}({W}_{m})\right]  ^{2}+\left[  \frac{J_{m}({W}_{m})}%
{\tan\theta}\right]  ^{2}\right\} \nonumber\\
&  +\frac{(m \tilde{\xi})\left(  q {\omega}_{m}\right)  ^{2} {\omega}_{m}}%
{\pi\gamma^{2}}\cos\theta\ \left[  J_{m}({W}_{m})\right]  ^{2}, \label{ar}%
\end{align}
in terms of the observed frequency $\omega_{m}=m \omega_{0}$. This
distribution shows a parity violation contribution proportional to
$\cos \theta$. The standard result for the angular distribution of
the spectral power given in Ref. \cite{SCHWBOOK} is directly
recovered when $\tilde{ \xi }=0$. Let us observe that the
expansion parameter $\tilde\xi$ appears amplified by the factor
$m$, which turns out to be very large in some astrophysical
settings, as we will discuss in section VIII.

\subsection{Total power radiated in the $m^{th}$-harmonic}

In the previous subsection we computed the unpolarized angular distribution of
the averaged power radiated into the $m^{th}$ harmonic, up to first order in
$\tilde{\xi}$. Such linear terms vanish when we integrate on the whole solid
angle, as it is evident by symmetry arguments. In this subsection we are
interested in the leading contribution to the unpolarized (angular) integrated
spectral averaged power distribution, which is of second order in
$z=\tilde{\xi}\omega$. We expect that such second order contribution will
still be amplified by factors having powers of $m$. In order to perform the
expansion let us introduce the notation
\begin{equation}
g(z)=\frac{n(z)}{\sqrt{1+z^{2}}}\left\{  2\beta^{2}J_{2m}^{\prime}\left[
2m\beta n(z)\right]  -\left[  n^{-2}(z)-\beta^{2}\right]  \int_{0}%
^{2mn(z)\beta}dxJ_{2m}(x)\right\}  , \label{GDEZ}%
\end{equation}
in such a way that we can write%
\begin{equation}
P_{m}=\frac{q^{2}m\omega_{0}}{2R}\left[  g(z_{m})+g(-z_{m})\right]
=P_{m,+}+P_{m,-}. \label{PME}%
\end{equation}
Expanding in powers of $z$ we have%
\begin{equation}
P_{m}=\frac{q^{2}m\omega_{0}}{2R}\left(  2g(0)+z_{m}^{2}g^{\prime\prime
}(0)\right)  . \label{PM}%
\end{equation}
The term $g(0)$ is readily obtained as
\begin{equation}
g(0)=\left\{  2\beta^{2}J_{2m}^{\prime}(2m\beta)-\gamma^{-2}\int_{0}^{2m\beta
}dxJ_{2m}(x)\right\}  . \label{GO}%
\end{equation}
The computation of $g^{\prime\prime}(0)$ is more laborious. It can
be written as
\begin{equation}
\left.  \frac{d^{2}g(z)}{dz^{2}}\right\vert _{z=0}=\left.  \frac{dg(n)}%
{dn}\right\vert _{n=1}+\left.  \frac{d^{2}g(n)}{dn^{2}}\right\vert _{n=1}.
\label{gz2}%
\end{equation}
Thus we need to calculate the first and second derivatives of $g(n)$ evaluated
at $n=1$. Our results are
\begin{align}
\left.  \frac{dg}{dn}\right\vert _{n=1}  &  =\left(  1+\beta^{2}\right)
\int_{0}^{2m\beta}dx\;J_{2m}(x)+2m\beta\left(  1-\beta^{2}\right)
J_{2m}(2m\beta),\label{DG}\\
\left.  \frac{d^{2}g}{dn^{2}}\right\vert _{n=1}  &  =-2\beta^{2}J_{2m}%
^{\prime}(2m\beta)-\left(  1+\beta^{2}\right)  \int_{0}^{2m\beta}%
dx\;J_{2m}(x)+\left(  \frac{2m\beta}{\gamma}\right)  ^{2}J_{2m}^{\prime
}(2m\beta),
\end{align}
which lead to
\begin{equation}
\left.  \frac{d^{2}g(z)}{dz^{2}}\right\vert _{z=0}=2m\beta\gamma^{-2}%
J_{2m}(2m\beta)+\left(  4m^{2}\gamma^{-2}-2\right)  \beta^{2}J_{2m}^{\prime
}(2m\beta\;). \label{DDG}%
\end{equation}

Going back to Eq. (\ref{PM}) giving the total power in the $m^{th}$ harmonic,
we obtain our final result
\begin{align}
P_{m}  &  =\frac{q^{2} \omega_{m}}{R}\left\{  2\beta^{2}J_{2m}^{\prime
}(2m\beta)-\gamma^{-2}\int_{0}^{2m\beta}dxJ_{2m}(x)\right. \nonumber\\
&  \left.  +\left(  \tilde{\xi} \omega_{m}\right)  ^{2}\beta\left[
\gamma^{-2}\left[  mJ_{2m}(2m\beta)+2m^{2}\beta J_{2m}^{\prime}(2m\beta
)\right]  -\beta J_{2m}^{\prime}(2m\beta)\right]  \right\}  . \label{pm1}%
\end{align}
When $\tilde{\xi}=0$ we recover the well established result for the total
power radiated in the m$^{th}$ harmonic \cite{SCHWBOOK}. Again we notice the
dominant amplifying factor $m^{2}$ in the fourth term in the RHS of Eq.
(\ref{pm1}).

\subsection{Polarization index}

Now we make an expansion to first order in $\tilde{\xi}$ of the polarization
index in Eq. (\ref{PINDEXP}). To this order
\begin{equation}
W_{m\lambda}\simeq\left(  1+\lambda\tilde{\xi}\omega\right)  m\beta\sin\theta,
\end{equation}
together with
\begin{align}
J_{m}(W_{m\lambda})  &  \simeq J_{m}(m\beta\sin\theta)+\lambda\tilde{\xi
}\omega m\beta\sin\theta\;J_{m}^{\prime}(m\beta\sin\theta),\label{JMPEXP}\\
J_{m}^{\prime}(W_{m\lambda})  &  \simeq J_{m}^{\prime}(m\beta\sin
\theta)-\lambda\tilde{\xi}\omega\left[  J_{m}^{\prime}(m\beta\sin
\theta)+m\frac{\beta^{2}\sin^{2}\theta-1}{\beta\sin\theta}J_{m}(m\beta
\sin\theta)\right]  , \label{JMP1EXP}%
\end{align}
where we have substituted the second derivative of the Bessel function arising
in Eq. (\ref{JMP1EXP}) via its differential equation. Substituting in Eq.
(\ref{PINDEXP}) we obtain
\begin{equation}
R_{m}=\left(  \frac{1-\tilde{\xi}\omega}{1+\tilde{\xi}\omega}\right)
\frac{\beta J_{m}^{\prime}(W_{m-})-\cot\theta J_{m}(W_{m-})}{\beta
J_{m}^{\prime}(W_{m+})+\cot\theta J_{m}(W_{m+})}. \label{RM}%
\end{equation}
The basic building blocks can be rewritten as%
\begin{align}
\beta J_{m}^{\prime}(W_{m-})-\cot\theta J_{m}(W_{m-})  &  =P_{-}+\tilde{\xi
}\omega Q_{-},\\
\beta J_{m}^{\prime}(W_{m+})+\cot\theta J_{m}(W_{m+})  &  =P_{+}-\tilde{\xi
}\omega Q_{+},
\end{align}
where%
\begin{align}
P_{-}  &  =\beta J_{m}^{\prime}(m\beta\sin\theta)-\cot\theta J_{m}(m\beta
\sin\theta),\label{PQMEN}\\
Q_{-}  &  =\beta(1+m\cos\theta)J_{m}^{\prime}(m\beta\sin\theta)+m\frac
{\beta^{2}\sin\theta^{2}-1}{\sin\theta}J_{m}(m\beta\sin\theta),\\
P_{+}  &  =\beta J_{m}^{\prime}(m\beta\sin\theta)+\cot\theta J_{m}(m\beta
\sin\theta),\\
Q_{+}  &  =\beta(1-m\cos\theta)J_{m}^{\prime}(m\beta\sin\theta)+m\frac
{\beta^{2}\sin\theta^{2}-1}{\sin\theta}J_{m}(m\beta\sin\theta). \label{PQMAS}%
\end{align}

In terms of the above quantities, and to first order in $\tilde{\xi}$, we
arrive at
\begin{equation}
R_{m}=\tilde{R}_{m}\left[  1-\tilde{\xi}\omega\left(  2-\frac{Q_{-}}{P_{-}%
}-\frac{Q_{+}}{P_{+}}\right)  \right]  . \label{RMPOL}%
\end{equation}
Here we have introduced
\begin{equation}
\tilde{R}_{m}=\frac{P_{-}}{P_{+}}=\frac{\beta J_{m}^{\prime}(m\beta\sin
\theta)-\cot\theta J_{m}(m\beta\sin\theta)}{\beta J_{m}^{\prime}(m\beta
\sin\theta)+\cot\theta J_{m}(m\beta\sin\theta)}, \label{RMTIL}%
\end{equation}
which corresponds to the polarization index in the case $\tilde{\xi}=0$. Using
Eqs. (\ref{PQMEN}-\ref{PQMAS}) we can evaluate the first order correction in
$\tilde{\xi}$ to $R_{m}$
\begin{equation}
2-\frac{Q_{-}}{P_{-}}-\frac{Q_{+}}{P_{+}}=\frac{2}{\sin^{2}\theta}\frac
{\cos^{2}\theta J_{m}(m\beta\sin\theta)-m\beta\gamma^{-2}\sin^{3}\theta
J_{m}^{\prime}(m\beta\sin\theta)}{\left[  \cot\theta J_{m}(m\beta\sin
\theta)\right]  ^{2}-\left[  \beta J_{m}^{\prime}(m\beta\sin\theta)\right]
^{2}}. \label{FACRMCOR}%
\end{equation}
The polarization index determines the Stokes parameters, given by Eqs.
(\ref{sp}). Using Eqs. (\ref{RM}), (\ref{RMTIL}) and (\ref{FACRMCOR}), we get
in particular%
\begin{equation}
u_{m}=-\tilde{\xi}\omega\frac{4\tilde{R}_{m}}{1+\tilde{R}_{m}^{2}}%
=-2\tilde{\xi}\omega\frac{\left[  \beta\sin\theta J_{m}^{\prime}(m\beta
\sin\theta)\right]  ^{2}-\left[  \cos\theta J_{m}(m\beta\sin\theta)\right]
^{2}}{\left[  \beta\sin\theta J_{m}^{\prime}(m\beta\sin\theta)\right]
^{2}+\left[  \cos\theta J_{m}(m\beta\sin\theta)\right]  ^{2}}, \label{um}%
\end{equation}
which gives a neat signature of the Lorentz violation.

\section{Contributions from dimension six operators}

Up to this point we have considered the Myers-Pospelov model as an exact one,
which allows us to make quantitative predictions of the ensuing effects. The
inclusion of higher dimension LIV operators in the action will certainly
modify the results obtained for the higher order contributions in the
corresponding parameters, i.e. $\tilde{\xi}^{n+1}$ and $\tilde{\eta}^{n+1}$
with $n>0$. The investigation of additional dimension six operators is an
interesting problem of its own, but it is beyond the scope of this paper and
we do not consider it in any detail here. Nevertheless, it is important to
emphasize that the contributions of these operators do not produce
birefringent effects, and thus this type of effects by dimension five
operators can be clearly disentangled from dimension six contributions by
polarization measurements, for example as indicated in Eq. (\ref{um}).

We can perform a rough estimation of the induced effects due to
the addition of dimension six operators in the Maxwell action for
the total power radiated into the $m^{th}$ harmonic, on the basis
of the following considerations. In the first place, since the
propagation of the electromagnetic field is basically
characterized by the refraction index, the addition of dimension
six operators leads to a modification dominated by a term
proportional to $\omega^{2}$
\begin{equation}
n(\lambda z)\longrightarrow\bar{n}(\lambda z)=n(\lambda z)+\tilde{\xi}%
_{2}\omega^{2},
\end{equation}
where $\tilde{\xi}_{2}=\xi_{2}/M^{2}$ is the parameter associated
to dimension six operators, and $n(\lambda z)$ is given by Eq.
(\ref{REFIND}). The total radiated power into the $m^{th}$
harmonic remains given by Eq. (149), but with modified $g(\lambda
z_{m})$ functions
\begin{equation}
g(\lambda z_{m})\longrightarrow\bar{g}(\lambda z_{m})=g(\bar{n}(\lambda
z_{m}))\;F(\lambda\omega_{m}),
\end{equation}
where the real function $F(\lambda\omega_{m})=F(\lambda\tilde{\xi}\omega
_{m},\tilde{\xi}_{2}\omega_{m}^{2})$ depends upon $m$ only through $\omega
_{m}$, and satisfies $F(\tilde{\xi}=\tilde{\xi}_{2}=0)=1$. Thus we have
\begin{equation}
P_{m}=\frac{q^{2}\omega_{m}}{2R}\left[  g(\bar{n}(z_{m}))F(\omega_{m}%
)+g(\bar{n}(-z_{m}))\;F(-\omega_{m})\right]  , \label{PMEE}%
\end{equation}
instead of Eq. (\ref{PME}). The corresponding expansions are now
\begin{equation}
F(\lambda\omega_{m})=1+A\lambda\tilde{\xi}\omega_{m}\;+\left(  B\;\tilde{\xi
}^{2}+C\;\tilde{\xi}_{2}\right)  \omega_{m}^{2}+....,
\end{equation}
where we assume that $A,B$ and $C$ are of order one together with
\begin{equation}
g(\bar{n}(z_{m}))=g(z_{m})+\tilde{\xi}_{2}\omega_{m}^{2}\left[  \frac
{dg(n)}{dn}\right]  _{\tilde{\xi}_{2}=0}+O(M^{-3}).
\end{equation}
Since we already have a factor $\tilde{\xi}_{2}\propto M^{-2}$ in front of the
above derivative we can further set there $\tilde{\xi}=0$, which means that it
is evaluated at $n=1$, and thus
\begin{equation}
g(\bar{n}(z_{m}))=g(z_{m})+\tilde{\xi}_{2}\omega_{m}^{2}\left[  \frac
{dg(n)}{dn}\right]  _{n=1}+O(M^{-3}).
\end{equation}
Replacing in Eq. (\ref{PMEE}) and recalling that $g(z)\simeq g(0)+zg^{\prime
}(0)+\frac{1}{2}z^{2}g^{\prime\prime}(0)$, together with $g^{\prime
}(0)=\left[  dg(n)/dn\right]  _{n=1}$, we finally obtain for the total
radiated power
\begin{equation}
P_{m}=\frac{q^{2}\omega_{m}}{2R}\left\{  2g(0)\left[  1+\left(  \tilde{\xi
}^{2}+\tilde{\xi}_{2}\right)  \omega_{m}^{2}\right]  +\left[  \tilde{\xi}%
^{2}g^{\prime\prime}(0)+2\left(  2\tilde{\xi}^{2}+\tilde{\xi}_{2}\right)
\left(  \frac{dg(n)}{dn}\right)  _{n=1}\right]  \omega_{m}^{2}\right\}  .
\end{equation}
The factor of $g(0)$ in the RHS shows corrections of order $\omega_{m}%
^{2}/M^{2}$ that we can safely neglect. The second square bracket also
contains contributions arising from dimension five and six operators. Using
Eqs. (\ref{GO}), (\ref{DG}) and (\ref{DDG}) the corresponding magnitudes can
be written as
\begin{align}
X_{m5}  &  =4\omega^{2}\tilde{\xi}^{2}\left\vert \left[  m^{2}\gamma
^{-2}J_{2m}^{\prime}(2m\beta)+\int_{0}^{2m\beta}dx\;J_{2m}(x)\right]
\right\vert ,\\
X_{m6}  &  =4\omega^{2}\left\vert \tilde{\xi}_{2}\left[  \int_{0}^{2m\beta
}dx\;J_{2m}(x)+m\gamma^{-2}J_{2m}(2m\beta)\right]  \right\vert .
\end{align}
Using the corresponding expressions for large $m$ given in Eqs. (\ref{BESSEXP}%
), and the further approximation (\ref{LAKNU1}) for the case $m/m_{c}%
\lesssim1$ we obtain
\begin{equation}
\frac{X_{m6}}{X_{m5}}=\frac{\left\vert \tilde{\xi}_{2}\right\vert \left[
1+0.36\;m^{2/3}\gamma^{-2}\right]  }{\tilde{\xi}^{2}\left[  1+0.26\;m^{4/3}%
\gamma^{-2}\right]  }.
\end{equation}
Thus, the effects of dimension six operators are negligible in the whole range
of the spectrum if $\left\vert \tilde{\xi}_{2}\right\vert /\tilde{\xi}^{2}<1$.
On the other hand, if $\left\vert \tilde{\xi}_{2}\right\vert /\tilde{\xi}%
^{2}\gtrsim\gamma^{2}$ they are significant even in the high
energy limit, near the cutoff frequency where
$m=m_{c}=3\gamma^{3}/2$. At the moment there is not enough
experimental evidence to set this point. For example, Ref.
\cite{GM} gives the boundaries $\left\vert \tilde{\xi}\right\vert
\lesssim10^{-12}$ and
$-10^{-6}\lesssim\tilde{\xi}_{2}\lesssim10^{-4}$. In the following
we restrict ourselves to the Myers-Pospelov model with no
admixture of dimension six operators.

\section{High energy limit}

In Table \ref{tab:a} we present a rough estimation of the relevant parameters
associated to some observed cosmological objects corresponding to SNR and GRB.
There $r\,[l.y.]$ is the distance of the object to the earth, $\gamma$ is the
Lorentz factor of the charged particles, $B\,[Gauss]$ is the average magnetic
field producing the synchrotron radiation, $\omega_{c}\,[GeV]$ is the cut-off
frequency and $\omega_{0}\,[GeV]$ is the Larmor frequency. In all cases the
cut-off frequency $\omega_{c}$ has been estimated from the radiation spectrum
fitted by a synchrotron model in the corresponding reference. This, together
with the magnetic field $B$ allows us to estimate the Lorentz factor
\begin{equation}
\gamma=2.36\times10^{8} \sqrt{\frac{\omega_{c}[GeV]}{B[Gauss]}\frac{M}{m_{e}}%
},
\end{equation}
where $M$ is the mass of the charged particle. From the above we further
obtain the Larmor frequency
\begin{equation}
\omega_{0}[GeV]=0.6 \times10^{-17}\left(  \frac{2m_{e}}{\gamma M} \right)
B[Gauss].
\end{equation}
In the case of Crab Nebula we adopt the typical values given in Ref.
\cite{ATOYAN}. For Mkn 501 we consider two possible models for synchrotron
radiation where the emitter particles are either protons \cite{AHARONIAN} or
electrons \cite{KONOPELKO}. In the latter case we use the radius of the orbit
$r^{\prime}=1/\omega_{0}=1.5 \times10^{15} \, cm$ and the magnetic field to
obtain the Lorentz factor. Finally we consider the GRB021206 ($z \approx1.25
$). According to Ref. \cite{NAKAR} this object has a very compact core with a
radius of the order of $1 \, km$ and a magnetic field $\approx10^{12} \;
Gauss$. The synchrotron emission region is about $10^{8} \, km$ from the core
\cite{WAXMAN}, so that we estimate the magnetic field to be $10^{4} \; Gauss$
using the transport law $B\,r = const.$. From Ref. \cite{HAJDAS} we take the
cut-off frequency as $\omega_{c}=1 \; MeV$.

As indicated in Table \ref{tab:a}, the radiation of interest is dominated by
very high harmonics $10^{15}\leq m \leq10^{30}$ exhibiting also high ratios of
$m/\gamma$, typically in the range $10^{10}\leq m/\gamma\leq10^{22}$. The
corresponding values for $\gamma$ imply also $\beta\approx1$. In this way, the
high harmonics present in the synchrotron radiation of these astrophysical
sources and the $\gamma$ factor of the corresponding radiating charges
highlight the relevance of the large $m$ and $\gamma$ limit with the
constraint $\left(  m/\gamma\right)  ^{2}\gg1$ to study Lorentz violating
effects. In the following we present the most interesting results of this
limit, from the phenomenological point of view.

\subsection{Large $m$ approximation}

We start from Eq. (\ref{PMDO}) by recalling the expressions for each helicity
contribution to the integrated power spectrum
\begin{align}
P_{m\lambda}  &  =\frac{q^{2} \omega_{m}}{R}\frac{1}{1+n^{2}(\lambda z_{m}%
)}\left\{  \frac{}{} 2\beta^{2}n^{2}(\lambda z_{m})J_{2m}^{\prime}\left[
2m\beta n(\lambda z_{m})\right]  \right. \nonumber\\
&  \qquad\qquad\qquad\qquad\quad\left.  -\left[  1-\beta^{2}n^{2}(\lambda
z_{m})\right]  \int_{0}^{2mn(\lambda z_{m})\beta}dx\;J_{2m}(x)\right\}  ,
\label{PMLI}%
\end{align}
where $n(\lambda z)$ are given in Eq. (\ref{REFIND}). According to Ref.
\cite{SCWANNP}, the important point here is the presence of two different
regimes of what is called synergic synchrotron-Cerenkov radiation, depending
upon $\left(  1-\beta^{2} n^{2}(\lambda z_{m})\right)  \lessgtr0$. Using
expressions (\ref{UMB2}) and (\ref{REFIND}), this basic combination can be
written in terms of the electron $E$ and photon $\omega_{m}$ energies as
\begin{align}
1-\beta^{2}n^{2}  &  =\left(  \mu^{2}/E^{2}\right)  \left(  1+2\omega_{m}%
^{2}\tilde{\xi}^{2}\right)  -2\omega_{m}^{2}\tilde{\xi}^{2}+2\tilde{\eta
}E-\frac{15}{4}\tilde{\eta}^{2}E^{2}-2\lambda\omega_{m} \tilde{\xi}\left[
1-2\tilde{\eta}E-\left(  \mu^{2}/E^{2}\right)  \right] \nonumber\\
&  + O(\tilde{\eta}^{2}\tilde{\xi})+O(\tilde{\eta}^{3})+O(\tilde{\xi}%
^{3})+O(\tilde{\eta}\tilde{\xi}^{2}), \label{1MB2N2}%
\end{align}
which contains the correction terms we are interested in.

Moreover, in our case we have%
\begin{equation}
\beta^{2}n^{2}(z)>\beta^{2}n^{2}(-z),
\end{equation}
so that we have three different possibilities for the total power
$P_{m}=P_{m+}+P_{m-}$: (i) $n^{2}(\lambda z)\beta^{2}<1$ for $\lambda=\pm1 $,
(ii) $n^{2}(\lambda z)\beta^{2}>1$ for $\lambda=\pm1$ and (iii) $n^{2}%
(+z)\beta^{2}>1$ and $n^{2}(-z)\beta^{2}<1$. They are combinations of two
possible situations. Let us recall the corresponding expressions of the
integrated power spectrum (\ref{PMLI}) for these basic situations.
\begin{table}[ptb]
\caption{Typical parameters of some relevant astrophysical objects}%
\label{tab:a}
\begin{ruledtabular}
\begin{tabular}{cccccccccccc}
&$r$&$\gamma$&$B$& $\omega_c$ &$\omega_0$& $m$ &$m/\gamma$\\
\hline CRAB& $10^4$ & $10^9$ &$10^{-3}$  &$10^{-1}$
& $10^{-30}$ &$10^{29}$& $10^{20}$  \\
${{\rm (Mkn\;501)}}_p$ & $10^{8}$ &$10^{11}$  &$10^{2}$ &$10^{4}$
& $10^{-29}$ & $10^{33}$ &$10^{22}$  \\
${{\rm (Mkn\;501)}}_e$& $10^{8}$ &$10^{11}$  &$10^{-1}$  &$10^{4}$
& $10^{-29}$ & $10^{33}$ &$10^{22}$ \\
GRB\;021206&$10^{10}$ & $10^{5}$ &$10^{4}$  &$10^{-3}$ &
$10^{-18}$& $10^{15}$&$10^{10}$
\\
\end{tabular}
\end{ruledtabular}
\end{table}

\subsubsection{Case $1 > \beta^{2}n^{2}\left(  \lambda z\right)  $.}

Introducing the notation
\begin{equation}
1-n^{2}\left(  \lambda z_{m} \right)  \beta^{2}=\left(  \frac{3}{2\tilde
{m}_{\lambda c}}\right)  ^{2/3}, \label{cm}%
\end{equation}
together with the large $m$ expressions given in Eq. (\ref{BESSEXP}) of
Appendix B we obtain%
\begin{equation}
P_{m\lambda}=\frac{q^{2} \omega_{m}}{\sqrt{3}\pi R\left(  1+n^{2}(\lambda
z_{m})\right)  }\left(  \frac{3}{2\tilde{m}_{\lambda c}}\right)  ^{2/3}\left[
\int_{m/\tilde{m}_{\lambda c}}^{\infty}dx\ K_{5/3}\left(  x\right)  -2\left(
\frac{3}{2\tilde{m}_{\lambda c}}\right)  ^{2/3}K_{2/3}\left(  m/\tilde
{m}_{\lambda c}\right)  \right]  , \label{TPMN1}%
\end{equation}
where the required values of $n=n(\lambda z_{m})$ should be substituted in the
sequel. Using the bremsstrahlung function defined in (\ref{BF}) and the
asymptotic limits given in Eqs. (\ref{LABF}), (\ref{LAKNU1}) and (\ref{LAKNU2}%
), we obtain
\begin{equation}
P_{m\lambda}=\frac{q^{2} \omega_{0} m^{1/2}}{2R\sqrt{\pi}\left(
1+n^{2}\right)  }\left(  \frac{3}{2\tilde{m}_{\lambda c}}\right)
^{1/6}\left[  1-2\left(  \frac{3}{2\tilde{m}_{\lambda c}}\right)
^{2/3}\right]  e^{-m/\tilde{m}_{\lambda c}}, \label{PHELE1}%
\end{equation}
for $m/\tilde{m}_{\lambda c}>>1$, and%
\begin{equation}
P_{m\lambda}=\frac{q^{2} \omega_{0} \left(  3\right)  ^{1/6}\Gamma(2/3)}{\pi
R\left(  1+n^{2}\right)  }\left[  1-\left(  \frac{3}{2\tilde{m}_{\lambda c}%
}\right)  ^{2/3}\right]  m^{1/3}, \label{PHELE2}%
\end{equation}
for $m/\tilde{m}_{\lambda c}<<1$. These expressions are similar
to the standard ones up to the factor $1/\left( 1+n^{2}\right)$
which might contribute to the corrections in the LIV parameters.
Equation (\ref{PHELE1}) exhibits $\tilde{m}_{\lambda c}$ as the
cut-off harmonic, which also depends on the LIV parameters.

\subsubsection{Case $1 < \beta^{2}n^{2}\left(  \lambda z\right)  $.}

Using the results of Eqs. (\ref{BESSEXP1}) in the Appendix we obtain the
expression analogous to Eq. (\ref{TPMN1})
\begin{equation}
P_{m}(n)=\frac{q^{2}m\omega_{0}}{R}\frac{(n^{2}\beta^{2}-1)}{1+n^{2}}\left[
\Lambda(x)-2\left(  n^{2}\beta^{2}-1\right)  \frac{Ai^{\prime}(-x)}{x}\right]
. \label{TPMN2}%
\end{equation}

As shown in \cite{SCWANNP} the expressions (\ref{TPMN1}) and (\ref{TPMN2}) are
continuous in $n\beta=1$. The Airy functions of negative argument involved in
this sector are oscillating functions in the large $x$ limit.

\subsection{Angular distribution of the $m^{th}$ harmonic to first order in
$\xi$}

To compute the unpolarized average angular distribution of the power radiated
into the $m^{\mathrm{th}}$ harmonic in the high energy limit we can start from
Eq. (\ref{ar}). By performing the required expansions we get
\begin{equation}
\frac{dP_{m}}{d\Omega}=\frac{\omega_{m}^{2}q^{2}}{2\pi}\left(  1-m\tilde{\xi
}\omega_{m} \cos\theta\right)  \left\{  \left[  J_{m}^{\prime}(m\beta
\sin\theta)\right]  ^{2}+ {\cot^{2}\theta}\ [J_{m}(m\beta\sin\theta
)]^{2}\right\}  . \label{EXPANG}%
\end{equation}
This expression shows an anisotropy, which is an effect of first order in
$\tilde{\xi}$. The emission is suppressed for $m\tilde{\xi}\omega_{m}
\cos\theta\simeq1$ and enhanced for $m\tilde{\xi}\omega_{m} \cos\theta\simeq-1
$. Note that both effects are amplified by the presence of factor
$m=\omega/\omega_{0}$ which can be very large in some regimes. When $m$ is
large, and since we have already expanded around ${\tilde\xi}$, we can use the
standard result that the radiation becomes confined to a small angular range
$\Delta\theta\simeq m^{-1/3}$ around $\theta=\pi/2$ \cite{SCHWBOOK}. Thus, if
we consider the radiation in the frontiers of the beam, i.e. $\theta\simeq
\pi/2\pm m^{-1/3}$, the anisotropy becomes significant when
\begin{equation}
\omega_{m} \simeq\left(  \omega_{0}^{2}/\tilde{\xi}^{3}\right)  ^{1/5}.
\label{fa}%
\end{equation}

\subsection{Total power radiated into the $m^{th}$ harmonic for large $m$ to
second order in $\tilde{\xi}$}

The total power radiated into the $m^{th}$ harmonic in the large $m$ limit can
be computed directly from Eq. (\ref{pm1}), which gives
\begin{align}
P_{m}  &  \simeq\frac{q^{2}\omega_{m}}{R}\left\{  2\beta^{2}J_{2m}^{\prime
}(2m\beta)-\gamma^{-2}\int_{0}^{2m\beta}dxJ_{2m}(x) +2\tilde{\xi}^{2} \left(
\frac{m\omega_{m}\beta}{\gamma}\right)  ^{2} J_{2m}^{\prime}(2m\beta)\right\}
, \label{PMHE}%
\end{align}
where we only keep the dominant term proportional to $m^{2}$ in the fourth
term of the RHS in Eq. (\ref{pm1}). Substituting in Eq. (\ref{PMHE}) the
asymptotic representations (\ref{BESSEXP}) for the Bessel functions for $n=1$,
with $m_{c}$ being the corresponding critical value
\begin{equation}
m_{c}=\frac{3}{2}\gamma^{3}, \label{MCRIT}%
\end{equation}
we get
\begin{align}
P_{m}  &  \simeq\frac{q^{2} \omega_{m}}{\sqrt{3}\pi R\gamma^{2}}\left[
\frac{m_{c}}{m}\kappa\left(  \frac{m}{m_{c}}\right)  -\frac{2}{\gamma^{2}%
}K_{2/3}\left(  \frac{m}{m_{c}}\right)  +2\tilde{\xi}^{2} \left(
\frac{m\omega_{m}\beta}{\gamma}\right)  ^{2} K_{2/3}\left(  \frac{m}{m_{c}%
}\right)  \right]  .
\end{align}
For large $m$ and $(1-\beta^{2})>0$ the behavior of the power radiated in the
$m^{th}$ harmonic can be obtained using the asymptotic expressions
(\ref{LABF}), (\ref{LAKNU1}) and (\ref{LAKNU2}). For the case $m/m_{c}\gg 1$ and
$\gamma\gg1$ the result is
\begin{equation}
P_{m}\simeq\frac{q^{2}\omega_{0}}{2\sqrt{\pi}R}\left(  \frac{m}{\gamma
}\right)  ^{1/2}\left[  1-\frac{2}{\gamma^{2}}+2\tilde{\xi}^{2} \left(
\frac{m\omega_{m}\beta}{\gamma}\right)  ^{2} \right]  e^{-2m/3\gamma^{3}}.
\end{equation}

Since the regime of interest has $m/\gamma\gg1$, we keep the contribution from
the LIV part that is amplified by the factor $m^{2}/\gamma^{2}$, obtaining
\begin{equation}
P_{m}\simeq\frac{1}{2}\left(  \frac{m}{\pi\gamma}\right)  ^{1/2}\;\left(
\frac{q^{2}B}{E}\right)  ^{2}\left(  1-3\tilde{\eta}E+\frac{27}{4}\tilde{\eta
}^{2}E^{2}\right)  \left[  1+2\tilde{\xi}^{2}\left(  \frac{m\omega_{m}}%
{\gamma}\right)  ^{2}\right]  e^{-2m/3\gamma^{3}}, \label{FINRES}%
\end{equation}
where we have rewritten $\omega_{0}/R$ in terms of the electron energy $E$
using (\ref{RADIOUS}) together with (\ref{OMEGA0}). We also set $\beta=1$ in
all factors where no divergence arises.

The complementary range $m/m_{c}\ll 1$ yields
\begin{equation}
P_{m}\simeq\frac{\sqrt[3]{9}\,\Gamma(2/3)}{\sqrt{3}\pi}\left(  \frac{q^{2}%
B}{E}\right)  ^{2}\left(  1-3\tilde{\eta}E+\frac{27}{4}\tilde{\eta}^{2}%
E^{2}\right)  \left[  1+\tilde{\xi}^{2} \left(  \frac{m\omega_{m}}{\gamma
}\right)  ^{2} \right]  m^{1/3}. \label{FINRES1}%
\end{equation}

In Eqs. (\ref{FINRES}) and (\ref{FINRES1}) the Lorentz factor $\gamma$ is
given either by Eq. (\ref{g1}) or (\ref{g2}), according to the value of
$\tilde{\eta}E^{3}/\mu^{2}$.

\subsection{Polarization in the large $m$ limit}

The general expression for the polarization index $R_{m}$, from which we can
calculate the reduced Stokes parameters, is given in Eq. (\ref{RMPOL}). Since
the radiation is mostly concentrated around the plane of the orbit, we will
examine the different limits of
\begin{equation}
R_{m}(\theta=\pi/2)=1-2\frac{\tilde{\xi}m\omega_{m} }{\beta\gamma^{2}}%
\frac{J_{m}(m\beta)}{J_{m}^{\prime}(m\beta)}.
\end{equation}
In the large $m$ approximation and in the case $1-\beta^{2}>0$, one can write
the Bessel function and its derivative in terms of the Macdonald functions as
stated in Eqs. (\ref{BESSEXP})\ of the Appendix. Using (\ref{LAKNU1}) and
(\ref{LAKNU2}) the latter can be further approximated according to $\frac
{m}{{m }_{c}}\lessgtr1$, where ${m}_{c}$ is given by Eq. (\ref{MCRIT}). When
$m/{m}_{c}\gg 1$ the power of the radiation is exponentially damped and we have
\begin{equation}
\frac{J_{m}(m\beta)}{J_{m}^{\prime}(m\beta)}=\gamma,
\end{equation}
which leads to
\begin{equation}
R_{m}(\theta=\pi/2)=1-\frac{2\tilde{\xi}\omega}{\beta}\frac{m}{\gamma}.
\end{equation}
On the other hand, when $m/{m}_{c}\ll 1$, we have
\begin{equation}
\frac{J{\,}_{m}(m\beta)}{J_{m}^{\prime}(m\beta)}=\left(  \frac{1}{4}\right)
^{1/3}\frac{\Gamma(1/3)}{\Gamma(2/3)}\, m^{1/3},%
\end{equation}
yielding
\begin{equation}
R_{m}(\theta=\pi/2)=1-\frac{2\tilde{\xi}\omega}{4^{1/3}\beta} \frac
{\Gamma(1/3)}{\Gamma(2/3)}\frac{m^{4/3}}{\gamma^{2}}.
\end{equation}
In both cases there appear amplifying factors such as $m$ and $m^{4/3}$, whose
effects have to be evaluated in each separate situation. Let us also remark
that the limit $\tilde{\xi}=0$ reproduces a linear polarization as dictated by
the value $R_{m}^{2}=1$.

\subsection{Averaged degree of circular polarization for large $m$}

We can often assume that the relativistic electrons $(\beta\simeq1)$ have an
energy distribution of the type \cite{RL}
\begin{equation}
N(E)dE=C\;E^{-p}\;dE, \label{EEDIST}%
\end{equation}
in some energy range $E_{1}<E<E_{2}$, where typically $p\simeq2-3$. Let us
define the circular analog of the degree of linear polarization introduced in
Ref. \cite{RL} as
\begin{equation}
\Pi_{\odot}=\frac{\left\langle P_{+}(\omega)-P_{-}(\omega)\right\rangle
}{\left\langle P_{+}(\omega)+P_{-}(\omega)\right\rangle }, \label{CIRCPOL}%
\end{equation}
where $P_{\pm}(\omega)$ is the total power distribution per unit frequency and
polarization $\lambda=\pm1$. The average $\left\langle ...\right\rangle $ on the
energy distribution of the charged particles is calculated as
\begin{equation}
\left\langle Q(E)\right\rangle =C\int_{0}^{\infty}E^{-p}\;Q(E)\;dE.
\end{equation}

In (\ref{PMDO}) we have introduced the power emitted into the $m^{th}$
harmonic $P_{m\lambda}$, such that the total power emitted with polarization
$\lambda$ is
\begin{equation}
P_{\lambda}=\sum_{m}P_{m\lambda}=\int P_{m\lambda}\;dm=\int\frac{P_{m\lambda}%
}{\omega_{0}}\;d\omega,
\end{equation}
which produces%
\begin{equation}
P_{\lambda}(\omega)=\frac{P_{m\lambda}}{\omega_{0}}.
\end{equation}
We will carry the calculation only to first order in the LIV parameters. To
this end we start from Eq. (\ref{TPMN1}) and set $\beta=1=n(\lambda z_{m})$
everywhere, except in the critical terms involving ${\tilde{m}}_{\lambda c}$
from where the corrections arise. We should also take into account that most
of the radiation comes from the terms with $m\approx{\tilde{m}}_{\lambda
c}\gg 1$, where $K_{2/3}(1)=0.49,\,\,{\kappa}(1)=0.65$. Thus the dominant term
is
\begin{equation}
P_{\lambda}(\omega)=D\left(  \frac{1}{\tilde{m}_{\lambda c}}\right)
^{2/3}\frac{\tilde{m}_{\lambda c}}{m}{\kappa}\left(  \frac{m}{\tilde
{m}_{\lambda c}}\right)  , \label{DOMTERM}%
\end{equation}
where $D$ is a constant. Starting from the definition (\ref{CIRCPOL}), making
an expansion in $u=\lambda{\tilde{\xi}}\omega$ and recalling that $\tilde
{m}_{\lambda c}=\tilde{m}_{\lambda c}(u)$, we find
\begin{equation}
\Pi_{\odot}={\tilde{\xi}}\omega\frac{\left\langle \left[  \frac{dP_{\lambda
}(\omega)}{du}\right]  _{{\tilde{\xi}}=0,\,\,{\tilde{\eta}}=0}\right\rangle
}{\left\langle \left[  P_{\lambda}(\omega)\right]  _{{\tilde{\xi}%
}=0,\,\,{\tilde{\eta}}=0}\right\rangle }+O({{\tilde{\xi}}}^{2},\;{\tilde{\xi}%
}{\tilde{\eta}},\dots). \label{CIRCPOL1}%
\end{equation}
Since $n(u)=\sqrt{1+u^{2}}+u$, we can rewrite $d/du$ in terms of $d/dn$.
Besides, $\tilde{m}_{\lambda c}$ depends only on the combination $n\beta$ so
that we can further go to $d/d\beta$ obtaining
\begin{equation}
\left.  \frac{dP_{\lambda}(\omega)}{du}\right|  _{{\tilde{\xi}}=0,\,\,{\tilde
{\eta}}=0}=D\left(  \frac{2n^{2}}{1+n^{2}}\frac{\beta}{n}\right)
_{{{\tilde{\xi}}=0,\,\,{\tilde{\eta}}=0}}\,\frac{d}{d\beta}\left[  \left(
\frac{1}{m_{c}}\right)  ^{2/3}\frac{m_{c}}{m}{\kappa}(m/m_{c})\right]  ,
\label{DER1}%
\end{equation}
where the first parenthesis on the RHS gives $1$ upon evaluation. In the
second parenthesis we have already taken the limit $n=1,{\tilde{\eta}}=0$. In
this case we also have $m_{c}=3\gamma^{3}/2$ and $E=\mu\gamma$. Therefore, the
argument of the bremsstrahlung function $\kappa(x)$ becomes
\begin{equation}
x=\frac{m}{m_{c}}=\frac{2}{3}\frac{\omega}{\omega_{0}}\gamma^{-3}%
=A^{2}\,\gamma^{-2}, \label{XVAR}%
\end{equation}
where $A^{2}=2\mu\omega/(3qB)$. The above leads to $\gamma=A\,x^{-1/2}$.
Again, a successive change of independent variables yields
\begin{equation}
d/d\beta=\beta\,\gamma^{3}\,d/d\gamma=-2A^{2}\,d/dx, \label{CHARUL}%
\end{equation}
where we have set $\beta=1$ in the corresponding factor. Finally, we have to
analyze the term $x\,m_{c}^{2/3} $ that multiplies\textbf{\ } $\kappa(x)$
inside the square bracket of Eqs. (\ref{DOMTERM}) and (\ref{DER1}).
Substituting $m_{c}=3\gamma^{3}/2=3\,A^{3}\,x^{-3/2}/2$ we find that
$x\,m_{c}^{2/3}=(3A^{3}/2)^{2/3}$ is a constant that cancels when taking the
ratio in (\ref{CIRCPOL1}). Thus we are left with
\begin{equation}
\Pi_{\odot}=-\frac{4}{3}{\tilde{\xi}}\omega\;\left(  \frac{\mu\omega}%
{qB}\right)  \,\frac{\left\langle \frac{d\kappa(x)}{dx}\right\rangle
}{\left\langle \kappa(x)\right\rangle }, \label{CPEXP1}%
\end{equation}
where the average $\left\langle ... \right\rangle $ over $E$ has been replaced by
one over $x$, via the change of variables $E=\mu\,A\,x^{-1/2}$, with all the
constant factors cancelling in the ratio of the two integrals. It should be
pointed out that the factor $A^{2}$ arising from the derivative in the
numerator survives after taking this ratio. For the purpose of making an
estimation of $\Pi(p)$ we take the energy range to be $\,0\leq E\leq\infty$.
We then obtain
\begin{equation}
\Pi_{\odot}=-\frac{4}{3}{\tilde{\xi}}\omega\;\left(  \frac{\mu\omega}%
{qB}\right)  \,\,\frac{\int_{0}^{\infty}x^{\left(  p-3\right)  /2}%
\frac{d{\kappa}(x)}{dx}\;dx}{\int_{0}^{\infty}x^{\left(  p-3\right)
/2}\;{\kappa}(x)\;dx}. \label{CPEXP2}%
\end{equation}
Using the expression \cite{RL}
\begin{equation}
\int_{0}^{\infty}x^{\mu}\;{\kappa}(x)\;dx=\frac{2^{\mu+1}}{\mu+2}\Gamma\left(
\frac{\mu}{2}+\frac{7}{3}\right)  \Gamma\left(  \frac{\mu}{2}+\frac{2}%
{3}\right)  ,\qquad\mu+1/3>-1,
\end{equation}
and comparing with Eq. (\ref{CPEXP}), we finally get
\begin{equation}
\Pi_{\odot}={\tilde{\xi}}\omega\left(  \frac{\mu\omega}{qB}\right)  \,\Pi(p),
\label{CPEXP}%
\end{equation}
with
\begin{equation}
\Pi(p)=\frac{(p-3)\left(  3p-1\right)  }{3\left(  3p-7\right)  }\,\frac
{(p+1)}{(p-1)}\frac{\Gamma\left(  \frac{p}{4}+\frac{13}{12}\right)
\Gamma\left(  \frac{p}{4}+\frac{5}{12}\right)  }{\Gamma\left(  \frac{p}%
{4}+\frac{19}{12}\right)  \Gamma\left(  \frac{p}{4}+\frac{11}{12}\right)
},\qquad p>7/3. \label{PIP}%
\end{equation}
The constraint $p>7/3$ is required to avoid the divergence of the
integral in the numerator of Eq. (\ref{CPEXP2}) at $x=0$. The
above result for $\Pi _{\odot}$ is the analogous for the averaged
linear degree of polarization $\Pi_{RL}=\left(  p+1\right) /\left(
p+7/3\right)$  which is independent of the frequency \cite{RL}. On
more realistic grounds, one should avoid the infinite upper limit
of $E$, (the zero lower limit of $x$). In this case the divergence
of the integral at $x=0,$ and hence the mathematical constraint on
$p$, disappears, but the expression for $\Pi(p)$ becomes more
complicated.

In any case, the most important feature of the result in Eq. (\ref{CPEXP}) is
the presence of the amplifying factor $(\mu\omega/qB)$, which is independent
of such details. An estimation of this factor in the zeroth-order
approximation (${\tilde{\xi}}=0={\tilde{\eta}}$), which is appropriate in Eq.
(\ref{CPEXP}), yields $(\mu\omega/qB)=\omega/(\omega_{0}\gamma)=m/\gamma$,
which is not necessarily a small number. According to the values that we have
mentioned at the beginning of this section, this amplifying factor could be as
large as $m/\gamma\simeq10^{22}$.

\section{Phenomenological consequences}

In our analysis of the synchrotron radiation we have found that a careful
perturbative expansion in the Lorentz violating parameters produced not only
the naively expected factors, but also non-trivial large amplifying factors
for some Lorentz violation effects. These unexpectedly large amplification
factors open the possibility of observing such effects in the radiation of
astrophysical sources where these factors become important. In this section we
explore different aspects of the synchrotron radiation from astrophysical
objects such as SNRs and GRBs and the constraints upon the Lorentz violating
parameters imposed by their measurement.

\subsection{Radiation angular distribution}

This subsection is devoted to the study of the preferred
directions together with the opening angles in the Lorentz
violating circularly polarized synchrotron radiation. The standard
Lorentz covariant results (see Ref. \cite{SCHWBOOK} for example)
are that the radiation is confined to the forward particle
direction with an opening angle $\delta\theta\simeq\;\gamma^{-1}
$. The recent bounds on the LIV parameters, derived from the
synchrotron radiation observed from the Crab nebula, heavily rest
upon extrapolating some of these results, based on the assumption
that the LIV violations should only slightly modify the standard
ones \cite{JACOBSON1}. This has been the subject of some
controversy in the literature \cite{CONTRO}. For this reason it is
relevant to provide some answers arising from a specific model
calculation in which the analysis of Refs. \cite{JACOBSON1} can be
embedded.

Our starting point is the angular distribution of the power emitted in the
$m^{th}$ harmonic with polarization $\lambda$ given in Eq. (\ref{pmlt1}),
which we rewrite here
\begin{equation}
\frac{dP_{m\lambda}}{d\Omega}=\frac{\omega_{0}^{2}q^{2}}{4\pi}\frac{1}%
{\sqrt{1+z_{m}^{2}}}\left[  \lambda m\beta n(\lambda z_{m})J_{m}^{\;\prime
}(W_{\lambda m})+\,m\cot\theta\,J_{m}(W_{\lambda m})\right]  ^{2}.
\end{equation}
Recalling that we are dealing with large values of $m$ it is convenient to use
the corresponding approximations in (\ref{BESSEXP}). By doing this Eq.
(\ref{pmlt1}) leads to
\begin{align}
\frac{dP_{m\lambda}}{d\Omega}  &  =\frac{\omega_{0}^{2}q^{2}}{12\pi^{3}}%
\frac{1}{\sqrt{1+z_{m}^{2}}}\left\{  m\lambda\beta n(\lambda z_{m})(1-\left(
\beta n(\lambda z_{m})\sin\theta\right)  ^{2})K_{2/3}\left[  \frac{m}%
{3}(1-\left(  \beta n(\lambda z_{m})\sin\theta\right)  ^{2})^{3/2}\right]
\right. \nonumber\\
&  \left.  +m\cot\theta(1-\left(  \beta n(\lambda z_{m})\sin\theta\right)
^{2})^{1/2}K_{1/3}\left[  \frac{m}{3}(1-\left(  \beta n(\lambda z_{m}%
)\sin\theta\right)  ^{2})^{3/2}\right]  \right\}  ^{2}.
\end{align}
Using further the approximation corresponding to $m/\tilde{m} _{c}\gg 1$, we can
make explicit the asymptotic behavior of the Macdonald functions which yields
\begin{align}
\frac{dP_{m\lambda}}{d\Omega}  &  =\frac{\omega_{0}^{2}q^{2}}{8\pi^{2}}%
\frac{m}{\sqrt{1+z_{m}^{2}}}e^{-2m\left[  1-\left(  \beta n(\lambda z_{m}%
)\sin\theta\right)  ^{2}\right]  ^{3/2}/3}\nonumber\\
&  \times\left[  \lambda\beta n(\lambda z_{m})(1-\left(  \beta n(\lambda
z_{m})\sin\theta\right)  ^{2})^{1/4}+\cot\theta\;(1-\left(  \beta n(\lambda
z_{m})\sin\theta\right)  ^{2})^{-1/4}\right]  ^{2}. \label{EXPLM}%
\end{align}
The above corresponds to the use of the approximation (\ref{LAKNU2}) of
Appendix B. In fact this expression turns out to be a reasonable approximation
for values of the Macdonald function argument as low as $\vartheta=0.2$
($m\approx1$). In this case it gives an error of the order of $15 \%$, which
is acceptable for our purposes.

To visualize the geometry of the radiation beam we can rewrite this angular
distribution in terms of the angle $\alpha=\pi/2-\theta$. First we estimate
the direction of the maximum radiated power in the $m-th$ harmonic given by
$\alpha_{max}$. Recalling the definition (\ref{cm}) for ${\tilde m}_{\lambda
c} $ and introducing the variables
\begin{equation}
\mu=\frac{m}{{\tilde m}_{\lambda c}}, \qquad y=\lambda\, \left(  \frac{2
{\tilde m}_{\lambda c}}{3}\right)  ^{1/3} \sin\alpha, \label{CHV}%
\end{equation}
we can write the extremum condition for (\ref{EXPLM}) as
\begin{equation}
3\mu y\left(  \left(  1+y^{2}\right)  ^{2}+y\left(  1+y^{2}\right)
^{3/2}\right)  -y\sqrt{1+y^{2}}-\left(  1+y^{2}\right)  -1\simeq0 .
\label{EXTR}%
\end{equation}
We assume the above equation to be valid for $\mu\lessgtr1$, for reasons to be
explained \textit{a posteriori}. Also, numerical estimations show that $y>1$
$(y<1)$ when $\mu<1$ $(\mu>1)$, respectively. These considerations lead to the
solutions
\begin{align}
&  \sin\alpha_{max}\,\simeq\,\alpha_{max}\simeq\lambda\, \left(
\frac{2{\tilde m}_{\lambda c} }{3m }\right)  ^{2/3}m^{-1/3},\qquad\mu\gg
1,\label{AMUMAY1}\\
&  \sin\alpha_{max}\,\simeq\,\alpha_{max}\simeq\lambda\; 0. 64\;
(2m_{c})^{-1/3}, \qquad\qquad\mu= 1,\label{AMUM}\\
&  \sin\alpha_{max}\,\simeq\, \alpha_{max}\simeq\lambda\; (2m)^{-1/3},
\;\qquad\qquad\qquad\mu\ll 1, \label{AMUMEN1}%
\end{align}
The result (\ref{AMUMEN1}) justifies the use of expression (\ref{EXPLM}) for
$\mu< 1$. The argument $\vartheta_{max}$ of the corresponding Macdonald
functions for $\alpha_{max}$ is
\begin{equation}
\vartheta_{max}= \frac{m}{3}\left[  \sin^{2}\alpha_{max}+ \left(  \frac
{3}{2{\tilde m}_{\lambda c}}\right)  ^{2/3}\cos^{2}\alpha_{max} \right]
^{3/2}_{\mu<< 1}\simeq\frac{1}{6}.
\end{equation}
As explained after Eq. (\ref{EXPLM}), we are still very close to the range
where we have considered the approximation (\ref{LAKNU2}) to be acceptable.

Finally we estimate the opening angle of the radiation $\delta\theta$, defined
with respect to the angle of maximum radiation. To do this we only consider
the exponential factor in (\ref{EXPLM}) and determine the cutoff angle
$\alpha_{c}$, defined to be the angle where the power decreases by a factor
$1/e$ with respect to its maximum value. This leads to the equation
\begin{equation}
\frac{2}{3}m\left(  1-\left(  \beta n\cos\alpha_{c}\right)  ^{2}\right)
^{3/2} =1+\frac{2}{3}m\left(  1-\left(  \beta n\cos\alpha_{max}\right)
^{2}\right)  ^{3/2}, \label{ALPHAC}%
\end{equation}
together with
\begin{equation}
\delta\theta=2(\alpha_{c}-\alpha_{max}).
\end{equation}
The radiation occurs mostly in the sector $\mu\leq1$, thus by solving
(\ref{ALPHAC}) to lowest order and using the expressions (\ref{AMUM}) and
(\ref{AMUMEN1}) we find
\begin{align}
&  \delta\theta= 0.98\, m^{-1/3}, \quad\mu<< 1,\nonumber\\
&  \delta\theta_{c}= 1.25 \, m_{c}^{-1/3}, \quad\mu= 1, \label{OPENING}%
\end{align}
which effectively means $\delta\theta\approx m^{-1/3}$.

When it is possible to ignore the photon contribution to the LIV
that enters only through $n(z_{m})$, that is to say when we can
set $n=1$ as it is the case of Ref. \cite{JACOBSON1}, we recover
the result
\[
\delta\theta_{c} \simeq\left(  1-\beta^{2}\right)  ^{1/2}\simeq\gamma^{-1},
\]
for the opening angle, which was one of the starting assumptions
in this reference.

\subsection{Birefringence}

One of the most evident features of this Lorentz violating electrodynamics is
directly related with the propagation of the electromagnetic field and
manifests itself as a vacuum birefringence effect, i.e. circular polarizations
propagate with different velocities
\begin{equation}
v_{\pm}^{-1}=\sqrt{1+\omega^{2}\tilde{\xi}^{2}}\pm\omega\tilde{\xi}.
\end{equation}
In principle this difference produces a shift in the arrival time of the
different polarizations radiated by a given source, but in fact this time
delay is very small to be observed even for a propagation in cosmological
distances. A more sensible manifestation of this difference in the propagation
velocities is the polarization of the field, which changes with the distance.

\subsection{Radiation anisotropy}

Another interesting effect produced by the Lorentz violation is
the anisotropy in the emitted radiation. According to Eq.
(\ref{fa}) this anisotropy becomes apparent for $\omega\sim\left(
\omega_{0}^{2}/\tilde{ \xi}^{3}\right) ^{1/5}$, where the
radiation is suppressed in the region $0\leq\theta<\pi/2$
($\pi/2<\theta\leq\pi$) and enhanced in $\pi/2<\theta\leq\pi$
($0\leq \theta<\pi/2$) for $\tilde{\xi}>0$ ($\tilde{ \xi}<0$),
respectively. This effect gives a clear signature of the Lorentz
violation, and can be used to set a bound for $\tilde{\xi}$. In
this respect, the most favorable test in laboratory conditions
could be provided by an electron accelerator such as LEP. There
the energy of the electrons is $E\simeq55\ GeV$, the radius of the
orbit is $R\simeq4.25\ km$, yielding the Larmor and a cut-off
frequencies of $\omega_{0}\simeq5\times10^{-20}\ GeV$ and
$\omega_{c}\simeq5\times10^{-5}\ GeV$ respectively. The absence of
asymmetry in the radiation up to the critical frequency should
imply $\left\vert \tilde{\xi}\right\vert \lesssim 10^{-6}\
GeV^{-1}$. In fact this is a very weak bound. A more stringent one
could be obtained from astrophysical systems such as GRBs,
provided that the electromagnetic radiation of these objects
actually corresponds to synchrotron radiation.

\subsection{Cutoff frequency}

The exact high energy limit for the integrated power spectrum is given in Eq.
(\ref{PHELE1}). At this level we can identify the modifications induced by the
Lorentz violating effects in the cutoff frequency through the term in the
exponential $e^{-\omega/\omega_{c}}$, where
\begin{equation}
\omega_{c}=\frac{3}{2}\omega_{0}\left(  1-n^{2}\beta^{2}\right)  ^{-3/2}.
\end{equation}
The corresponding factors expressed in terms of the electron energy, given by
Eqs. (\ref{OMEGA0}) and (\ref{1MB2N2}), are rewritten here to first order in
the small dimensionless parameters\ $\xi,\eta$ previously introduced
\begin{align}
\omega_{0}  &  =\frac{|q|B}{E}\left(  1+\frac{3}{2}\eta\frac{E}{M}\right)  ,\\
1-\beta^{2}n^{2}  &  =\frac{\mu^{2}}{E^{2}}-2\eta\frac{E}{M}-2\lambda\xi
\frac{\omega}{M}.
\end{align}
Since the terms in $\left(  1-\beta^{2}n^{2}\right)  ^{-3/2}$ are already very
large it is appropriate to neglect the very small correction appearing in
$\omega_{0}$. In this way we have
\begin{equation}
\omega_{c}=\frac{3}{2}\frac{|q|B}{E}\left(  \frac{\mu^{2}}{E^{2}}-2\eta
\frac{E}{M}-2\lambda\xi\frac{\omega}{M}\right)  ^{-3/2}. \label{OMEGAC}%
\end{equation}
The relation of the above result with that given in Eq. (4) of
Ref. \cite{JACOBSON1} can be easily found. Changing ${\vec v}/c$
by ${\vec v}/c(\omega)$ in our definition (\ref{LORFACT}) of the
Lorentz factor we get
\begin{equation}
\gamma_{J}=\left\{  1-\left[  v(E)/c(\omega)\right]  ^{2}\right\}
^{-1/2}=\left(  1-n^{2}\beta^{2}\right)  ^{-1/2},
\end{equation}
where we have introduced the index of refraction $n=c/c(\omega)$. Let us
rewrite%
\begin{equation}
\gamma_{J}=c(\omega)\left\{  \left[  c(\omega)-v(E)\right]  \left[
c(\omega)+v(E)\right]  \right\}  ^{-1/2}=\left\{  2\left[  c(\omega
)-v(E)\right]  \right\}  ^{-1/2}%
\end{equation}
to the leading order. The dispersion relations (\ref{DR}) give
\begin{equation}
2\left(  c(\omega)-v(E)\right)  =-2\lambda\xi\frac{\omega}{M}+\frac{\mu^{2}%
}{E^{2}}-2\eta\frac{E}{M},
\end{equation}
in such a way that Eq. (\ref{OMEGAC}) can be written as
\begin{equation}
\omega_{c}=\frac{3}{2}\frac{|q|B}{E}\gamma_{J}^{3},
\end{equation}
which is precisely Eq. (4) of Ref. \cite{JACOBSON1}, up to the
polarization factor multiplying the photon frequency which has
been subsequently taken into account in later publications.
Nevertheless, in the considered energy range the photon
contribution is negligible and the following maximum energy is
obtained
\begin{equation}
E_{\max}=\left(  -\frac{2}{5\eta}\mu^{2}M\right)  ^{1/3}.
\end{equation}
It is possible to verify that
\begin{equation}
\left(  1-\beta^{2}n^{2}\right)  _{E=E_{\max}}=\frac{9}{5}\frac{\mu^{2}%
}{E_{\max}^{2}}>0,
\end{equation}
so that one stays in the allowed range of radiation emission.

\subsection{Additional corrections}

Another factor that modifies the power spectrum and tests the dynamics of the
electromagnetic field, according to Eq. (\ref{FINRES}), is $1+2\left(
\tilde{\xi}m^{2}\omega_{0}/\gamma\right)  ^{2}$. If this correction is not
observed at a given frequency $\omega$ we infer that $\tilde{\xi}\omega
^{2}/\left(  \omega_{0}\gamma\right)  \ll1$, or equivalently
\begin{equation}
\tilde{\xi}\lesssim\gamma\omega_{0}/\omega^{2}=qB/(\mu\omega^{2}). \label{xi}%
\end{equation}
This bound is independent of the energy of the electrons. It is proportional
to $\omega_{0}/\omega^{2}$, as in the parity violating effect already
discussed after Eq. (\ref{EXPANG}), but now we have an additional factor of
$\gamma$, which increases the resulting upper limit, thus making the boundary
much less stringent.

It is interesting to compare the bounds for $\tilde{\xi}$ provided by the
first order anisotropic effect in (\ref{FINRES}) and the second order
suppression in the power spectrum in (\ref{EXPANG}). According to Eq.
(\ref{fa}), the first case requires
\begin{equation}
\tilde{\xi}\lesssim\left(  \omega_{0}^{2}/\omega^{5}\right)  ^{1/3}=\left(
\omega/\omega_{0}\right)  ^{1/3}\omega_{0}/\omega^{2}, \label{BB1}%
\end{equation}
while in the second case Eq. (\ref{xi}) demands
\begin{equation}
\tilde{\xi}\lesssim\gamma\omega_{0}/\omega^{2}.
\end{equation}
Assuming that the Lorentz violation is not observed up to a frequency
$\omega<\omega_{c}=\gamma^{3}\omega_{0}$, the boundary (\ref{BB1}) leads to
\begin{equation}
\tilde{\xi}\lesssim\left(  \omega/\omega_{0}\right)  ^{1/3} \omega_{0}%
/\omega^{2}=\left(  \omega/\omega_{c}\right)  ^{1/3}\left(  \gamma\omega
_{0}/\omega^{2}\right)  <\gamma\omega_{0}/\omega^{2}.
\end{equation}
If $\omega\simeq\omega_{c}$ both relations (\ref{xi}) and
(\ref{BB1}) give the same bound for $\tilde{\xi}$.

Finally, Eq. (\ref{FINRES}) contains two additional frequency independent
factors including $\eta-$dependent corrections: $\left(  1+\tilde{\eta}%
E^{3}/2\mu^{2}\right)  $ that comes from $\gamma^{-1/2} $ through Eq.
(\ref{g1}) and $\left(  1-3\tilde{\eta}E+27\tilde{\eta} ^{2}E^{2}/4\right)  $.
Both test the dynamics of the charge. The first one dominates the linear
corrections at energies that we can find in astropysical systems, and if they
are not observed for electrons of energy $E$ this implies
\begin{equation}
\left\vert \tilde{\eta}\right\vert \lesssim\mu^{2}/E^{3}. \label{k1}%
\end{equation}

\section{Accessible parameter regions and general outlook}

To close our analysis, we will determine the regions of LIV parameters that
can be explored by means of the synchrotron radiation of the already
considered astrophysical sources: Crab Nebula, Mkn 501 and GRB021206. The Crab
Nebula is a well known synchrotron radiation emitter, while the other two are
on a more conjectural level.

Let us first consider the two LIV contributions to the power
spectrum Eq. (\ref{FINRES}) discussed in section IX-E. The one
containing $\tilde\eta$ is a frequency independent factor which
simply renormalizes the power spectrum. Since we do not have an
independent determination of the density of the emitter particles
we are not able to evaluate it. The second one, which contains
$\xi$ and depends on the frequency, distorts the shape of the
power spectrum thus being in principle measurable. If this
distortion is not observed in the power spectrum we can set the
bound (\ref{xi}). In the particular case of the CRAB Nebulae this
leads to $\xi<5\times10^{-1}$. Other strong sources of high energy
electromagnetic radiation are Mkn 501 and GRB 021206. Eq.
(\ref{xi}) shows that these objects could allow us to explore the
region $\xi\gtrsim10^{-6}$.

The values of the different factors contributing to the Green funcion phase
(\ref{PHASE}) for the astrophysical objects under consideration are given in
Table {\ref{tab:a1}}. Here we take the Larmor radius as an estimation of the
size of the radiating source $r^{\prime}$. The last column gives the
subdominant term in the phase of the Green function, $(r^{\prime}/r)^{2}$,
that allows us to estimate if a given LIV contribution is significant for the
radiation field, according to the discussion in Section III. In the Crab
Nebulae case Table {\ref{tab:a1}} tells us that ${\tilde{\xi}}\omega=\xi
\omega/M_{P}\ll (r^{\prime}/r)^{2}$, even for $\xi\approx\, 1$. In this way,
consistency requires that the phase of the Green function (\ref{PHASE})
reduces to $\omega\left(  r-\mathbf{\hat{n}}\cdot\mathbf{r^{\prime}}\right)
$. This is equivalent to set ${\tilde{\xi}}=0$, $n=1$ in all the arguments of
the Bessel functions. This leads to
\begin{equation}
\delta\theta\approx{\gamma}^{-1}(E),\qquad\omega_{c}=qB\gamma^{3}(E)/E,
\end{equation}
where $\gamma(E)$ still includes corrections depending upon the
fermionic parameter ${\tilde{\eta}}$ according to Eq.
(\ref{UMB2}). In others words, the results in Ref.
\cite{JACOBSON1} are recovered for this particular situation. On
the other hand, the corresponding phase in the cases of Mkn 501
and the GRB 021206, leading to modified expressions for the
cut-off frequency could, in principle, give information about
first order contributions in $\tilde\xi$. Since these corrections
are not affected by significant amplifying factors, their
observation would require an extremely precise determination of
the cut-off frequency. \begin{table}[ptb] \caption{Factors
contributing to the Green function phase for the
astrophysical objects}%
\label{tab:a1}
\begin{ruledtabular}
\begin{tabular}{cccccccccccc}
&$r'/r $&$\omega_c/M_P$&$(r'/r)(\omega_c/M_P)$
&$\left(r'/r\right)^2$ \\
\hline CRAB&$10^{-6}$&$10^{-20}$& $10^{-26}$&$10^{-12}$ \\
${{\rm Mkn\;501}}$&$10^{-11}$&$10^{-15}$&$10^{-26}$& $10^{-22}$ \\
GRB\;021206&$10^{-24}$&$10^{-22}$&$10^{-46}$&$10^{-48}$\\
\end{tabular}
\end{ruledtabular}
\end{table}
In this analysis we have considered the case of a charged particle
in a circular orbit orthogonal to the magnetic field, and we have
used our results to obtain some rough estimations of the regions
where the LIV parameters can be explored. In realistic systems
such as SNRs or GRBs, we actually have a population of charged
particles with a certain energy distribution and different pitch
angles. Thus, in order to derive more reliable bounds it is
necessary to take into account these distributions and perform the
corresponding averages. This process involves a rather detailed
model for the astrophysical source and is beyond the scope of the
present article.

To summarize, we have presented a complete analysis of the
synchrotron radiation in the rest frame of the Myers-Pospelov
model for a charged particle in circular motion perpendicular to a
constant magnetic field. The process can be visualized as a
standard electrodynamics radiation in a dispersive parity
violating media, with helicity-dependent refraction indices that
encode the Lorentz violating corrections to the electromagnetic
sector. The charged particle sector also presents corrections
arising from the Lorentz violation, which mainly modify the
dependence of the $\beta$ factor upon the energy of the particle.
The calculation includes exact expressions for the angular
distribution of the power (\ref{pmlt1}) and the total power
(\ref{PMDO}) radiated into the $m^{th}$ harmonic for each
polarization, together with their corresponding expansions to
lowest order in the LIV parameter $\tilde\xi$ given in Eqs.
(\ref{ar}) and (\ref{pm1}), respectively. The physical situations
under consideration correspond to large Lorentz factors $\gamma$
together with harmonics in the range $m \approx 10^{15}-10^{30}$,
but in such away that the ratio $m/\gamma$ is very large also. Due
to this fact, a large $m$ expansion has been performed in the
total power radiated into the $m^{th}$ harmonic for the exact
polarized case, Eqs. (\ref{PHELE1}-\ref{PHELE2}), and in the
unpolarized $\tilde\xi$-expanded situation, Eqs. (\ref{FINRES}
-\ref{FINRES1}). The $\tilde\xi$-exact case allows us to identify
the cutoff frequency according to Eq. (\ref{cm}). Special emphasis
has been given to the polarization of the radiation which exact
Stokes parameters are obtained in terms of Eq. (\ref{PINDEXP}),
with their corresponding expansion written in Eq. (\ref{RMPOL}).
We have also calculated the averaged degree of circular
polarization for a standard electron energy distribution in the
large $m$ approximation in Eq. (\ref{CPEXP}). This quantity starts
linearly in ${\tilde\xi}$ which, unexpectedly, is multiplied by
the amplifying factor $(\mu\omega)/(qB)\approx m/\gamma$. Such
amplifying factors occur also in other $\tilde\xi$-expanded
quantities. Similar amplifying factors have been obtained in
calculations of the synchrotron radiation spectra in the context
of non-commutative electrodynamics \cite{Castorina}. A study of
the observational relevance of these factors is outside the scope
of the present work, but certainly it is a matter that deserves
further investigation. As in the standard case, the radiation in
the large $m$ limit is concentrated in the forward direction.
Finally, the angles for maximum radiation together with the
corresponding opening angles are calculated in Eqs.
(\ref{AMUMAY1}), (\ref{AMUM}), (\ref{AMUMEN1}) and
(\ref{OPENING}), respectively.

\section*{Acknowledgements}

LFU acknowledges partial support from projects CONACYT-M\'exico-40745-F and
DGAPA-UNAM-IN104503-3; as well as the hospitality of the George P. and Cynthia
W. Mitchell Institute for Fundamental Physics, Texas A\&M University. RM
acknowledges partial support from CONICET-Argentina.

\appendix

\section{Circular polarization basis}

Given a wave propagating in the direction $\mathbf{k}$ we define the $\pm$
polarization components of a vector $\mathbf{j}=\{j_{i}\,(\omega,\mathbf{k}
)\}$ as
\begin{equation}
\mathbf{j}^{\pm}=\frac{1}{2}\left[  \mathbf{j}-\left(  \mathbf{\hat{k}}%
\cdot\mathbf{j}\right)  \mathbf{\hat{k}}\,\pm i(\mathbf{\hat{k}}%
\times\mathbf{j})\right]  , \label{JPM}%
\end{equation}
where we have introduced the unit vector $\mathbf{\hat{k}}=\mathbf{k}/|
\mathbf{k}|$. These components satisfy the following relations
\begin{align}
\mathbf{j}  &  =j\mathbf{^{+}}+j\mathbf{^{-}},\;\;\qquad\mathbf{j^{\pm}}%
\cdot\mathbf{l^{\pm}}=0\\
\mathbf{\hat{k}}\times\mathbf{j}^{+}  &  =-i\mathbf{j}^{+},\qquad
\qquad\mathbf{\hat{k}}\times\mathbf{j}^{-}=i\mathbf{j}^{-}. \label{RELJPM}%
\end{align}
with $\mathbf{l}$ being another arbitrary vector.

Sometimes it is convenient to rewrite (\ref{JPM}) in terms of projectors
$P_{ik}^{\pm}$ defined as
\begin{equation}
j_{i}^{\pm}=P_{ik}^{\pm}j_{k}\,,\qquad P_{ik}^{\pm}=\frac{1}{2}\left(
\delta_{ik}-{\hat{k}}_{i}{\hat{k}}_{k}\pm i\epsilon_{ijk}{\hat{k}}_{j}\right)
. \label{PROY}%
\end{equation}
The algebra of these projectors is very useful and is summarized in the
relations
\begin{align}
P_{ik}^{+}  &  =P_{ki}^{-}\,,\label{AP1}\\
P_{ki}^{\pm}\,P_{il}^{\pm}  &  =P_{kl}^{\pm}\,,\qquad\;\;\;\;\;\;P_{ki}^{\mp
}\,P_{il}^{\pm}=0\,,\label{AP2}\\
\epsilon_{pqr}\,{\hat{k}}_{p}\,P_{qs}^{\pm}\,P_{rt}^{\pm}  &
=0\,,\;\;\;\;\;\epsilon_{pqr}\,{\hat{k}}_{p}\,P_{qs}^{\pm}\,P_{rt}^{\mp}=\mp
iP_{st}^{-}\,. \label{AP3}%
\end{align}
Let us state that in the frame we have previously chosen we have the following
basis
\begin{equation}
\mathbf{\hat{k}=(}\sin\theta,0,\cos\theta),\;\;\;\;\mathbf{e}_{\parallel
}=(0,1,0),\;\;\;\;\mathbf{e}_{\perp}=(-\cos\theta,0,\sin\theta),
\end{equation}
with the associated circular basis
\begin{equation}
\mathbf{e}_{\pm}=\frac{1}{\sqrt{2}}\left(  \mathbf{e}_{\parallel}\pm
i\mathbf{e}_{\perp}\right)  ,\;\;\;\;\left(  \mathbf{e}_{\pm}\right)  ^{\ast
}\cdot\mathbf{e}_{\pm}=1,\;\;\left(  \mathbf{e}_{\pm}\right)  ^{\ast}%
\cdot\mathbf{e}_{\mp}=0.
\end{equation}
Thus a direct calculation shows that for an arbitrary complex current
$\mathbf{j=(}j_{x},j_{y},j_{z}\mathbf{)}$ the corresponding expressions for
the currents defined in Eq. (\ref{JPM}) are
\begin{equation}
\mathbf{j}^{\pm}=\frac{1}{\sqrt{2}}\left[  j_{y}\pm i\left(  \;j_{x}\cos
\theta-j_{z}\sin\theta\right)  \right]  \mathbf{e}_{\pm}. \label{PROYC}%
\end{equation}

\section{Large $m$ expansion of some required Bessel functions}

Using the Nicholson-Olber \cite{N0} asymptotic representations for the Bessel
functions in terms of the Macdonald functions $K_{n/m}$, we can write, in the
large $m$ approximation \cite{EWL,SCWANNP}:

\subsection{The case $1-n^{2}\beta^{2}>0$}

\begin{equation}
(1-n^{2}\beta^{2})=\left(  \frac{2}{3}\tilde{m}_{c}\right)  ^{-2/3}=\left(
\frac{3}{2\tilde{m}_{c}}\right)  ^{2/3},
\end{equation}%
\begin{align}
J{\,}_{2m}(2mn\beta)  &  \simeq\frac{1}{\sqrt{3}\pi}\left(  \frac{3}%
{2\tilde{m}_{c}}\right)  ^{1/3}K_{1/3}\left(  \frac{m}{\tilde{m}_{c}}\right)
,\nonumber\\
\;J{\,}_{2m}^{^{\prime}}(2mn\beta)  &  \simeq\frac{1}{\sqrt{3}\pi}\left(
\frac{3}{2\tilde{m}_{c}}\right)  ^{2/3}K_{2/3}\left(  \frac{m}{\tilde{m}_{c}%
}\right)  ,\nonumber\\
\int_{0}^{2m\;n\beta}dxJ_{2m}(x)  &  \simeq\frac{1}{\sqrt{3}\pi}\int
_{m/\tilde{m}_{c}}^{\infty}dxK_{1/3}\left(  x\right)  , \label{BESSEXP}%
\end{align}
where the cutoff number $\tilde{m}_{c}$ signals different regimes in the
behavior of the Bessel functions $K_{\nu}$ of imaginary argument.

A useful relation among the Macdonald functions is
\begin{equation}
2K_{2/3}^{\;\prime}\left(  x\right)  =-\left[  K_{5/3}\left(  x\right)
+K_{1/3}\left(  x\right)  \right]  .
\end{equation}

The bremsstrahlung function $\kappa(z)$ is defined as
\begin{equation}
\kappa(z)=z\int_{z}^{\infty}dxK_{5/3}\left(  x\right)  , \label{BF}%
\end{equation}
and has the following asymptotic behavior \cite{ERBER},
\begin{align}
\kappa(z)  &  \simeq2^{2/3}\Gamma(2/3)\;z^{1/3},\;\;\;z\ll 1,\nonumber\\
\kappa(z)  &  \simeq\sqrt{\frac{\pi}{2}}\;z^{1/2}\;e^{-z},\;\;\;\;\;\;z\gg 1.
\label{LABF}%
\end{align}
We also need the asymptotic behavior of $K_{\nu}(z)$
\begin{align}
K_{\nu}(z)  &  \simeq\frac{\Gamma(\nu)}{2}\left(  \frac{z}{2}\;\right)
^{-\nu},\;\ \ \ \ \;\;z\ll 1,\label{LAKNU1}\\
K_{\nu}(z)  &  \simeq\sqrt{\frac{\pi}{2}}\;z^{-1/2}\;e^{-z},\;\;\;\;\;\;z\gg 1.
\label{LAKNU2}%
\end{align}
where the last relation is independent of $\nu$.

\subsection{The case $1-n^{2}\beta^{2}<0$}

Here the notation is
\begin{equation}
x=m^{2/3}(n^{2}\beta^{2}-1),
\end{equation}
and we have
\begin{align}
J{\,}_{2m}^{^{\prime}}(2m\;n\beta)  &  \simeq-m^{-2/3}Ai^{\prime
}(-x),\nonumber\\
\int_{0}^{2m\;n\beta}J_{2m}(t)dt  &  =\frac{1}{3}+\int_{0}^{x}%
dt\;Ai(-t)\nonumber\\
\Lambda(x)  &  =-\frac{2Ai^{\prime}(-x)}{x}+\;\frac{1}{3}+\int_{0}%
^{x}dt\;Ai(-t), \label{BESSEXP1}%
\end{align}
where we have introduced the analogous $\Lambda(x)$ of the bremsstralung
function\ $\kappa(x)$.

\end{document}